\documentclass[letterpaper,twocolumn,10pt]{article}
\usepackage{usenix-2020-09}

\usepackage{ifpdf}
\usepackage{cite}

\usepackage{tikz}
\usepackage{amsmath}

\usepackage{arabtex}
\usepackage{utf8}
\usepackage[T1]{fontenc}
\usepackage[utf8]{inputenc}
\usepackage[main=english]{babel}

\usepackage{xspace} 
\usepackage{balance}
\usepackage{enumitem}
\usepackage{mathptmx}
\usepackage{graphicx}
\usepackage{array}
\newcolumntype{H}{>{\setbox0=\hbox\bgroup}c<{\egroup}@{}}
\usepackage{multirow}
\usepackage{booktabs}
\usepackage{tabularx}
\usepackage{hyperref}
\usepackage[strings]{underscore}
\usepackage{array}
\usepackage{soul}
\usepackage{subcaption}
\usepackage{xspace}
\usepackage{xparse}
\usepackage{amsmath}
\usepackage{anyfontsize}
\usepackage{color}
\usepackage{rotating}
\usepackage{amssymb}
\usepackage{adjustbox}
\usepackage{xcolor}
\usepackage{colortbl}
\usepackage{microtype}
\usepackage{cite}
\usepackage{tipx}
\usepackage{dblfloatfix}
\usepackage{etoolbox}
\usepackage{tikz}
\usepackage{floatpag}

\newrobustcmd*{\mycircle}[1]{\tikz{\filldraw[draw=#1,fill=#1] (0,0) circle [radius=0.1cm];}}

\newrobustcmd*{\mytriangle}[1]{\tikz{\filldraw[draw=#1,fill=#1] (0,0) --
(0.2cm,0) -- (0.1cm,0.2cm);}}

\definecolor{armygreen}{rgb}{0.0, 0.5, 0.0}

\newlength{\qswidth}
\newlength{\nswidth}

\colorlet{pastelgreen}{green!20!}
\colorlet{pastelred}{red!20!}
\colorlet{pastelyellow}{yellow!20!}
\definecolor{LightGray}{rgb}{0.88,0.87,0.88}

\sethlcolor{pastelgreen}

\newcommand{\hlc}[2][pastelgreen]{ {\sethlcolor{#1}\hl{#2}} }

\newcommand{\comment}[1]{\relax}
\newcommand{\eg}{e.g.\@\xspace}
\newcommand{\ie}{i.e.\@\xspace}
\newcommand{\etal}{et~al.\@\xspace}
\newcommand{\etc}{{\em etc}\xspace}

\newcommand{\expert}{high expertise\xspace}
\newcommand{\Expert}{High expertise\xspace}
\newcommand{\moderate}{moderate expertise\xspace}
\newcommand{\Moderate}{Moderate expertise\xspace}
\newcommand{\notomild}{limited expertise\xspace}
\newcommand{\Notomild}{Limited expertise\xspace}
\newcommand{\redacted}{Consumer Reports\xspace}

\newcommand*\emptycirc[1][0.5ex]{\tikz\draw (0,0) circle (#1);\xspace} 
\newcommand*\emptysquare[1][1ex]{\tikz\draw (0,0) rectangle (#1,#1);\xspace}

\makeatletter
\renewenvironment{quotation}
{\list{}{\listparindent=0pt
\itemindent    \listparindent
\leftmargin=10pt
\rightmargin=10pt
\topsep=1pt
\parsep        \z@ \@plus\p@}
\item\relax}
{\endlist}
\makeatother

\makeatletter
\renewenvironment{quote}
{\list{}{\listparindent=0pt
\itemindent    \listparindent
\leftmargin=4pt
\rightmargin=0pt
\topsep=0pt
\parsep        \z@ \@plus\p@}
\item\relax}
{\endlist}
\makeatother

\begin{document}

\date{}

\title{\Large \bf ``All of them claim to be the best'': Multi-perspective study of \\VPN users and VPN providers}

 \author{
 {\rm Reethika Ramesh}\\
 University of Michigan
 \and
{\rm Anjali Vyas}\\
 Cornell Tech
 \and
 {\rm Roya Ensafi}\\
 University of Michigan
 } 

\clubpenalty=10000\widowpenalty=10000
\maketitle

\subsection*{Abstract}  
As more users adopt VPNs for a variety of reasons, it is important to develop empirical knowledge of their needs and mental models of what a VPN offers. Moreover, studying VPN users alone is not enough because, by using a VPN, a user essentially transfers trust, say from their network provider, onto the VPN provider. To that end, we are the first to study the VPN ecosystem from both the users' and the providers' perspectives. In this paper, we conduct a quantitative survey of 1,252 VPN users in the U.S. and qualitative interviews of nine providers to answer several research questions regarding the motivations, needs, threat model, and mental model of users, and the key challenges and insights from VPN providers. We create novel insights by augmenting our multi-perspective results, and highlight cases where the user and provider perspectives are misaligned. Alarmingly, we find that users rely on and trust VPN review sites, but VPN providers shed light on how these sites are mostly motivated by money. Worryingly, we find that users have flawed mental models about the protection VPNs provide, and about data collected by VPNs. We present actionable recommendations for technologists and security and privacy advocates by identifying potential areas on which to focus efforts and improve the VPN ecosystem.

\section{Introduction}
\label{sec:intro}

Since their introduction over two decades ago, the use of Virtual Private Network (VPN) technologies has grown rapidly. With commercialization, VPN products have found their way into a regular Internet user's toolbox~\cite{cloudwards-vpn,infosecurity-vpnuse}.  Though the VPN ecosystem has expanded into a multi-billion dollar industry~\cite{prnewswire-vpn}, questions regarding why VPNs have been adopted so widely are still unanswered. Is the popularity of VPNs grounded in an understanding of risks from the users' part? Is the rise of VPNs due to dwindling trust in Internet service providers? What benefits do users perceive to gain?

A majority of previous studies have found various issues in the technical implementations of VPNs~\cite{ikram2016analysis,khan2018empirical,weinberg-imc18,vpnalyzer-ndss, xue2022OpenVPN}. Only limited prior work has delved into the human factors of VPN use: factors that contribute to retention of VPNs~\cite{namara2020emotional, zou2020examining}, attitudes of university students and corporate users towards VPNs~\cite{binkhorst2022security, dutkowska2020practicing, marshini-usenix-vpn}, and the widespread misconceptions of how privacy-enhancing tools work~\cite{story2021awareness}. 

However, no study has combined both the users and VPN providers perspectives to answer fundamental questions about the VPN ecosystem. For instance, users using VPNs are essentially transferring trust from their network provider onto the VPN provider, but it is unclear as to what VPN features encourages them to make this shift? On the other hand, the VPN industry has been known to employ various marketing tactics~\cite{akgulinvestigating} and dark patterns around discounts~\cite{dark-patterns,mathur2019dark}, but it is yet unknown if these practices are bound to have any significant effect on VPN users. Moreover, the community has not yet understood VPN providers' incentives in sustaining such dark patterns, nor do we know what efforts they take to foster user confidence in an ecosystem plagued with mistrust. To gain a clearer picture of the inner workings of such a large consumer ecosystem, it is imperative to study both its users and its providers.

This is the first multi-perspective study that uses a \textit{quantitative survey of \textbf{(n=1,252)} VPN users} in the U.S. along with \textit{qualitative interviews of nine leading VPN providers}. We choose to survey 1,252 users, that have either used or currently use a VPN, to provide us with practical insights into our various lines of inquiry that we systematize into the following research questions:

\textit{\textbf{RQ1: [Motivations]}} Why do users use VPNs?

\textit{\textbf{RQ2: [Needs and Considerations]}} What factors around VPNs do users consider when choosing a provider?

\textit{\textbf{RQ3: [Emotional Connection and Threat Model]}} How safe do users feel when browsing the internet with and without a VPN? (If and) From whom do users want to secure/conceal their online activity?

\textit{\textbf{RQ4: [Mental Model]}} Do users have an accurate understanding of how VPNs work and what data they collect?

\textit{\textbf{RQ5: [Perception and Trust]}} How do users perceive the VPN ecosystem?

\textit{\textbf{RQ6: [Alignment between VPN users and providers]}} What are the key areas of (mis)alignment in priorities and incentives between the two?

\begin{figure}[t] 
 \centering
 \includegraphics[width=0.99\columnwidth]{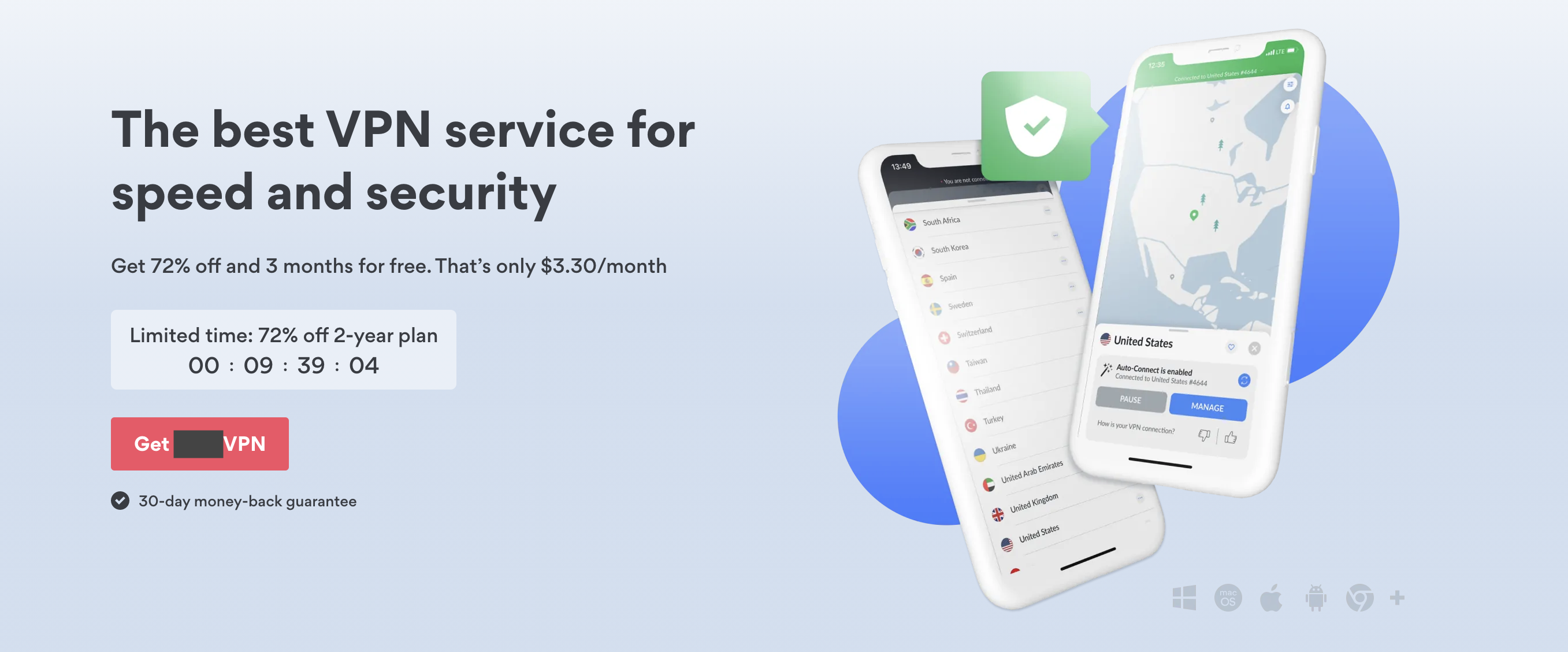}
 \caption{Example of dark pattern--using countdown timers.$\clubsuit$}
\label{fig:discount}
\end{figure}

We find that users rate speed, price, and easily understandable GUI, as the top requirements from VPNs rather than features such as the variety, number of available VPN servers, and their locations. We also find that in alignment with VPN providers' expectations, pricing plays a key role with users. Thus indicating that discounts, and marketing around pricing can have a significant effect on them. Prior research suggests that malicious marketing tactics~\cite{akgulinvestigating} and dark patterns around discounts are common, which are often used to ensure customer lock-in~\cite{dark-patterns,mathur2019dark}; an example of such a dark pattern in the VPN ecosystem is shown in Figure~\ref{fig:discount}.

Interestingly, we find that when it comes to choosing VPNs to use, more users seem to lean towards using search engines (61.1\%), and recommendation websites (56.5\%), rather than relying on more traditional methods such as word of mouth (5.7\%). Furthermore, almost 94\% of these users rate these websites trustworthy. On the other hand, our interviews with VPN providers highlights that the VPN recommendation ecosystem is mostly money motivated, with widespread malicious practices that include having paid review spots, and auctioning off the \#1 spot. Some ``review'' sites have been reported to send emails to VPN providers asking for higher cost-per-action/click to get ranked on their list~\cite{windscribe-review}. Users' reliance on such websites further amplifies our worries of an unregulated marketing ecosystem around VPNs.

Exploring reasons for why users use VPNs, we discover that users attach an emotional connection with using a VPN, namely a feeling of safety (86.7\%), which was found by prior work to be a key factor in retention of VPN use~\cite{namara2020emotional}. Our intuition suggested that exploring users' threat models could explain why they attach such considerations; indeed, we find that 91.5\% of users indicate they use VPNs for securing or protecting their online activity. When exploring who they aim to protect it from, we find that their top concerns are hackers/eavesdroppers on open WiFi networks (83.9\%), advertising companies (65.4\%), and internet service providers (46.9\%). This marks a departure from known prior concerns such as government surveillance (30\%), and indicates a shift of attitude towards surveillance capitalism and user privacy.

Given the emotional attachments and user dependency on VPNs for security and privacy concerns, we find that an alarmingly high proportion of users (39.9\%) have a \textit{flawed mental model} of what VPNs provide them and what data they collect. These users believe their ISP can still see the websites they visit over the VPN. More worryingly, we do not see significant difference between users of different expertise having flawed mental models ($\chi^{2}$-test, $p$=0.0927, N=1252). From our VPN provider interviews, we find that providers also mention that they recognize the need for improving user knowledge, and consider effective education a key challenge. We also find that dark patterns in the industry may also be a key issue; multiple VPN providers mention ``malicious marketing'' is problematic, including preying upon users' lack of knowledge and overselling of service.

Continuing to explore the confusion surrounding the operations of VPN providers, we find that a significant portion of \notomild users believe that the data is being collected for monetization, such as advertising (36.4\%), user tracking (36.4\%), and selling to third parties (33.6\%). Although a majority of all users (79.2\%) believe the main reason for data collection is internal analytics, confusion found amongst the \notomild users may be even more widespread among the VPN users in the general public. Moreover, users also expressed high degrees of concern towards VPN providers selling their data (73.2\%). This is yet another area of misalignment between users and providers, because multiple VPN providers believe that they clearly communicate their logging practices, and/or have released audits to prove this.

From our study of 1,252 users and 9 VPN providers, we present the following actionable recommendations for the VPN ecosystem: prioritizing user education, oversight on advertisements and marketing surrounding VPNs, coordinated efforts to bring attention to the flawed VPN recommendation ecosystem, and regulations to curb malicious marketing tactics that lead to false mental models and false expectations for users. We believe that our work will help security and privacy advocates such as EFF and CDT, technologists, and VPN providers alike, by calling attention to the key areas in the commercial VPN ecosystem.
\section{Background \& Related Work}
\label{sec:background}
\subsection{Virtual Private Networks}
\label{subsec:VPN}
Virtual Private Networks were initially created in 1996~\cite{vpn-origin} as a peer-to-peer tunnelling protocol developed in Microsoft to facilitate private communication in enterprise settings. Virtual private networks (VPNs) provided a way to create private connections between computers and transfer data between them securely over the public Internet. These are still the guarantees that VPNs provide for general users today. VPN products (VPNs hereon) create a secure connection, often called a ``tunnel'', to a secure server that then connects them to their intended destination. This tunnel typically provides an extra layer of encryption that serves as protection from surveillance by the intermediate networks, bypasses access restrictions active in those networks, and hides the user's actual IP address from their destination service~\cite{vpn-nytimes}.

Commercial VPN providers make use of the available VPN protocols such as OpenVPN, L2TP, IPSec, IKEv2, and Wireguard~\cite{openvpn,townsley1999layer,kaufman2005internet,donenfeld2017wireguard}, or develop proprietary protocols which are typically extensions of existing ones, optimized to fit their particular needs and business model. VPNs offer different subscription models: paid/premium services, free to use services, and freemium models that offer limited free features and charge for premium features and services. 

While some work has focused on analyzing technical aspects~\cite{vpnalyzer-ndss, tolley2021blind, khan2018empirical, ikram2016analysis}, Weinberg \etal~\cite{weinberg-imc18} focused on evaluating the claims of VPN server locations, and found at least one-third of the 2269 servers were definitely not in the country advertised, and another one-third probably were not in the location they claim. Investigating an often overlooked source of security advice, Akgul \etal~\cite{akgulinvestigating} studied 243 YouTube videos containing VPN ads and find a number of concerning misleading claims, including over-promises and exaggerations which may lead to users forming inaccurate mental models of internet safety. There have also been news reports of data breaches, leaks and misuse by VPN providers, some of which were published on VPN recommendation websites~\cite{malwarebytes-breach, vpnmentor-breach,techcruch-breach,cnet-misuse}.

\subsection{User Adoption of VPNs and Other Tools}
\label{subsec:user-adopt}

As users adopt more privacy-enhancing tools such as VPNs for a variety of reasons, their privacy needs become important to assimilate. The level of security and privacy a user needs may depend on myriad factors like the reasons for use, tolerance of failure, legality of these VPN services in the country of the user \etc. Only few community efforts focus on providing threat model based VPN (and other tools) recommendations, such as the Security Planner~\cite{security-planner}. We present the related work summarized in relevance to the topics studied:

\textbf{Prior work studying VPN users: }
Namara \etal~\cite{namara2020emotional} conducted a study with 90 technologically savvy users and studied the adoption and usage of VPNs, and the barriers they encounter in adopting them. They find users with emotional reasons to use a VPN such as fear of surveillance or desire for privacy, are more likely to continue using them rather users who use it for practical reasons. Similarly exploring the factors that influence user decisions to adopt VPN apps, Sombatruang \etal~\cite{sombatruang2020attributes} interviewed 32 users in UK and Japan and found that user review rating and price significantly influenced the choosing of a VPN to use.

\textbf{Prior work exploring particular sub-populations: }
Binkhorst \etal~\cite{binkhorst2022security} studied the mental models of 18 users in the context
of corporate VPNs, and found that experts and non-expert users have similar mental models of VPNs, and experts also tend to have false beliefs on security aspects of VPNs. Dutkowska-Zuk \etal conducted a study focused on a specific sub-population of 349 university students to find how and why they use VPNs, and whether they understand the various privacy risks caused by VPNs~\cite{dutkowska2020practicing}. They found that students are mostly concerned with access to content rather than privacy, and that most students did not use VPNs regularly. Extending this study, they looked at how these students compare to general VPN users in the awareness of risks of VPN use and how they adopt VPNs~\cite{marshini-usenix-vpn}. Specifically, they found that despite having different use cases, both groups had low understanding of the risks of data collection by VPNs, highlighting the need for better awareness campaigns.

\textbf{Prior work studied user attitudes and use of privacy-enhancing tools: }
Various prior works have shown that users, particularly in the U.S. are aware of risks such as tracking, and are concerned about online tracking in different situations~\cite{melicher2016preferences, chanchary2015user, rader2014awareness}. However, some prior work has shown that they are unclear on how to protect themselves~\cite{shirazi2014deters}. Story \etal~\cite{story2021awareness} highlighted this in their study of the use of and perceptions about web-browsing privacy related tools. In their survey of 500 U.S. users, they ascertain user perception of the protection provided by different tools across 12 different scenarios, and interestingly, they find that users having more experience using VPNs is associated with confusion about their protection. Further, studying the adoption and abandonment of 30 commonly recommended security and privacy practices, Zou \etal~\cite{zou2020examining} surveyed 902 users and find that security practices were more widely adopted and privacy related practices were among the ones most commonly abandoned.

These prior work, though useful, are limited in the scale, topics studied, and have focused on particular sub-populations of VPN users. \textbf{In our work, we create a novel line of inquiry to study the motivations, needs, and considerations of VPN users in depth, and improve greatly upon the scale of users surveyed. We are also the first to conduct a study of VPN providers.} We augment insights from both users and providers to characterize any misalignments between them, which could be exploited by bad actors to further deepen problems in the VPN ecosystem. Given the wide reach of the VPN ecosystem, our study will help technologists and security and privacy advocates gain a deeper understanding of the key problem areas where they can focus their efforts.
\section{Methods}
\label{sec:methods}

We set out to study VPN users and providers to understand their unique perspectives on the VPN ecosystem and the issues surrounding it. We conduct a large-scale survey of VPN users as well as a qualitative interview of nine VPN providers.

\subsection{User Survey}
\label{subsec:user-survey-methods}
\paragraph{\textbf{Small-Scale Interviews and Interactions.}}
We believe that a successful large-scale quantitative study must be preceded by a smaller-scale qualitative study and community research to extract key concerns. To that end, we conduct seven user interviews (4 men, 3 women, ages 18-45), and we participated in various VPN-focused community events with VPN providers and users in attendance in order to gather topics and research questions that interest the community.

For the small-scale user interview study, we design a questionnaire with open-ended questions to serve as the framework for each interview, and obtain approval from our Institutional Review Board (IRB). During the interview, we collect general demographic information, including gender, age range, occupation, country of residence, and level of education. Our introductory questions ask about the interviewee's awareness of their own threat model and of online risks such as trackers. Next, we ask about the perceived positives and negatives of VPN use. We then ask participants to sketch their understanding of how VPNs work while walking through the steps of setup and use, diagrammatically. The interview concludes with questions about how the VPN ecosystem can improve. 

We recruit seven participants via a pre-interview survey at the Citizen Lab Summer Institute~\cite{clsi-2019} that has global attendees who are passionate about technologies aiding Internet freedom, security, and user rights. Prior to the start of each interview, we obtain explicit consent for participation and permission to audio record it using an IRB-approved consent form. Participants are also given the chance to ask any questions before the interview begins and are allowed to stop at any point. After completion of the interviews, the first author transcribed all the recordings. Overall, the interviews lasted 15-20 minutes not including setup and conclusion.

\begin{table}[t]
    \centering
    \scriptsize
    \begin{tabular}{p{2cm}p{5.5cm}}
    \toprule
        \textbf{Themes} & \textbf{Definitions} \\
        \midrule
     	Reasons for using VPN &	Motivations for and reasons to use a VPN \\
     	\rowcolor{LightGray}
     	VPN Use &	General thoughts about commercial VPNs, what they look for \\
        Threat model for using a VPN &	Personal threat models for needing, using, and/or recommending a VPN \\
        \rowcolor{LightGray}
     	Mental Model of VPN & What is a VPN and what does it provide me? (Sketching exercise included) \\
     	Attitudes towards VPN services & What is lacking in current ecosystem, their perception of what the VPN ecosystem looks like \\
     	\rowcolor{LightGray}
     	Improving ecosystem & Thoughts to improve ecosystem and boosting adoption and safe usage \\
    	\bottomrule
    \end{tabular}
    \caption{Six themes with their definitions. $\clubsuit$}
    \label{tab:codes-themes}
\end{table}

\paragraph{\textbf{Developing the Large-Scale Survey Instrument.}}
After completing the interviews, we use an inductive open-coding method for analysis. Two members independently coded all the transcripts, and held a meeting to resolve any disagreements and create a codebook. The research team then met to collaboratively go through the codebook and identify emerging themes~\cite{braun2006using} and hence, we do not present inter-rater reliability for this case~\cite{mcdonald2019reliability,miles2014qualitative}. We augment these with the knowledge extracted from attending several Internet freedom community gatherings organized around VPNs and VPN use including the IFF VPN Village~\cite{iffvv-2020}. Finally, we combine our work to arrive at six common themes, shown in Table~\ref{tab:codes-themes}.

Using these themes, we devise an initial survey instrument to study VPN users. The instrument contains questions aimed at understanding users' motivations, needs, and considerations when it comes to VPN services, and discerning their threat models, perceptions of VPNs, and understanding of how VPNs work. During the design phase, we also create a consent form and obtain IRB approval. Our survey questions only collect the information we need and do not involve the collection of any personally identifiable information. 

\paragraph{\textbf{Cognitive Pre-testing.}}
In order to reduce the potential for biases that arise from the ordering and/or phrasing of questions, we conduct systematic pretesting in \textit{three phases}, iteratively improving the survey between each phase. 

First, we recruit test participants (from the target demographic, VPN users) at an Internet freedom, security and privacy focused event organized by the Open Tech Fund~\cite{otf} to pretest the initial survey instrument and obtain unbiased opinions about the survey. The pretesting involves vocally stepping through the survey while a facilitator from our team takes notes. We use these notes to detect biases, signs of confusion regarding the intent of the question, as well as ``leading'' questions. In this round, 17 pretesters worked through the survey, and we learned that comparison-scale adjectives (None at all, Little, Somewhat) were unclear for participants. We amend the scales to avoid ambiguity and provided clearer distinctions \eg we use Likert-type Scale when asking about concern or importance. The scale is provided on the Qualtrics software and is a psychometric scale developed for scaling responses in survey research~\cite{likert1932technique}. We learned that participants had varying understandings of what ``commercial VPNs'' mean, and that participants were not sure if the questions pertained to personal or professional VPN use. To remove ambiguity, we define ``commercial VPNs'' on the survey landing page and present examples within the survey.

After refining our survey using the initial pretesting, we requested external user-study experts to go through our survey and provide feedback. They helped us refine our matrix style questions and simplify the organization of our survey. 

After incorporating expert feedback, we run the last round with eight new pretesters. This round helped us refine some of the examples used in the survey, improve consistency of language, and disambiguate a handful of questions.

\paragraph{\textbf{The Final Survey Instrument.}}
The final survey instrument contains six parts, 28 questions (with sub-parts), and we incorporate one quality check (where they must confirm they use a VPN) and two attention checks. The survey starts with a demographic section, where we follow the community best practices for inclusive language, and also have a ``prefer not to disclose'' option for all demographic questions~\cite{spiel2019better, redmiles2017summary}. Then, we ask users general questions about their VPN usage, reasons for using a VPN, the resources used in discovering VPNs, importance of different criteria and features, their mental model of VPNs and the data it collects, their emotional connections tied to their use of a VPN (\eg safety), and their expectations from a VPN provider. We specifically avoid using words such as privacy and security in the text, since these concepts are broad, subjective, and mean different things to different users. Instead we allow users to select from list of options, and we distill into certain buckets during analysis. The final survey instrument is available in the Appendix~\ref{app:user-survey}.

\paragraph{\textbf{Analysis of Survey Data.}}
For the quantitative data from the survey, we report the results summarizing the users' responses. We aim to understand how different subgroups of users answer the same question, \ie users with different security and privacy expertise, and users who prefer to use certain subscription type (free or paid VPNs). We conduct $\chi^{2}$-tests, where all the assumptions are satisfied in each case, to examine if users in different subgroups of the same type (\eg expertise) answer questions differently. If there were significant differences between subgroups, we conduct pairwise comparison Z-tests ($\alpha$=0.05), where we adjust the significance levels for multiple comparisons through the FDR-BH adjustment~\cite{benjamini1995controlling} and present how they compare to each other.

We analyze the survey participants' open-ended text box responses using inductive coding. A primary coder created an initial codebook and assigned codes to all responses. A second coder analyzed 20\% of the responses for each coded question and ensure high inter-rater reliability~\cite{landis1977measurement}. Cohen's $\kappa$ between the two raters is 0.81, 0.86, 0.75, 0.81 for each question in Appendix~\ref{app:codes}, indicating moderate to strong agreement~\cite{cohen1960coefficient,mchugh2012interrater}.
The coders also coded responses for ``Other'' write-in options in questions, and present the responses in the results (\S\ref{sec:results}).

\subsection{Qualitative Interviews of VPN Providers}
\paragraph{\textbf{Interview Instrument.}}
Using the same parent themes as mentioned in Table~\ref{tab:codes-themes}, we create a questionnaire to interview VPN providers. These questions aim to extract insights from the providers about their users, VPN users in general, their business decisions, and what they see as the main issues in the VPN ecosystem. We design the topics for the questions to be counterparts to the VPN user survey.

\paragraph{\textbf{Interview Procedure.}}
We design a semi-structured interview with eight broad open-ended questions, and five additional questions to ask in case we have time. We obtain IRB approval prior to conducting the interviews. Our questionnaire, presented in Appendix~\ref{app:provider-questionnaire}, serves as a framework for the interviews to ensure we maintain structure and consistency from one provider to another. However, we also explore statements made by the interviewees for clarity and insights.

We begin all the interviews by presenting the interviewees with an overview of our project. Using an IRB-approved consent form, we obtain explicit consent for participation and audio recording the interview. On average, the interviews were $\approx$44 minutes in length, not including set up and conclusion. We conclude the interviews by thanking the participants, and provide ways to contact us to learn more about our project.

\paragraph{Analysis using Qualitative Coding.}
The first author transcribed all the interviews and the analysis is done using inductive open-coding, and thematic analysis~\cite{braun2006using}. Although we have nine VPN providers, we have eight transcripts in total because two of the providers opted to interview together\footnote{They are (non-commercial) partner projects, with separate services.} and each of them answered each question independently. A primary coder coded all transcripts, and two additional coders independently coded five and three transcripts each. Then, the team went over each coded transcript together to reconcile any differences. We then collaboratively identify the emerging themes for each question, and common themes that appear across different questions. Since the team collaboratively analyzed the coded transcripts together to identify themes, we do not present inter-rater reliability~\cite{mcdonald2019reliability,miles2014qualitative}.

Since this interview is meant to shed light on the VPN provider's perspectives and form a clearer picture of the VPN ecosystem, we only report aggregate results after performing thematic analysis. We anonymize the comments and do not attribute statements to particular providers.

\subsection{Recruitment}
\label{subsec:recruitment}

\paragraph{User Survey.}
In partnership with \redacted, a leading consumer research and advocacy organization with over 6 million members, we launched our user survey on March 1, 2021. We ask VPN users to participate in our survey by distributing the recruitment message in \redacted' tech-focused mailing list, subreddits such as \texttt{r/VPN}, \texttt{r/asknetsec}, \texttt{r/samplesize}, and on Twitter using the research team's own personal accounts. We also request participation from users on mailing lists belonging to Open Tech Fund and Internet Freedom Festival. We opt to recruit participants organically and to ensure anonymity, we did not offer any compensation for taking the survey. 

\paragraph{VPN Provider Interviews.}
We reached out to 15 leading VPN providers; nine of whom agreed to our interview. We chose to contact commercial VPN providers based on their popularity in the U.S., and included non-commercial projects that develop VPNs for users, based on their involvement in the Internet freedom and anti-surveillance community. We did not compensate the interviewees for participation.

The VPN providers we interviewed are the following (in alphabetical order): CalyxVPN, Hide.me, IVPN, Jigsaw Outline, Mullvad VPN, RiseupVPN, Surfshark, TunnelBear VPN, and Windscribe. We interviewed CEOs, CMOs, and/or researchers working in the company who were authorized to speak to us on behalf of the company.

\subsection{Ethics}
Our user study is approved as exempt from ongoing review under Exemption 2 as determined by our Institutional Review Board (IRB), and the VPN provider interview received a ``Not Regulated'' status. Furthermore, we draft a privacy policy document that was reviewed by experts from \redacted, and add it to our website. We also provide information on our study's Qualtrics page and ensure that our participants, pretesters, and interviewees explicitly consent to the study.

We follow user survey best practices such as using mindful, inclusive language in collecting demographics data~\cite{spiel2019better}. We also offer ``prefer not to answer'' as an option on our required demographics questions as per American Association for Public Opinion Research code of ethics~\cite{lazar2017research, redmiles2017summary}. We did not collect any personally identifiable information from our participants, and our results from the VPN providers are anonymized as well.

We solicit participation as mentioned in \S\ref{subsec:recruitment} and to ensure anonymity, we offer no compensation for any of our studies. We deeply analyze the collected responses to ensure response quality, as we detail in \S\ref{sec:data}. Audio-recordings of the interviews (both the small-scale user ones, and the VPN providers) were only accessed by the first author who did all the transcriptions. 

\subsection{Limitations}

As with many user surveys, some of our comparisons rely on self-reported data, which is prone to biases. We take efforts to reduce these biases to our best extent, elaborated in \S\ref{sec:data}, such as by explicitly explaining the different levels of privacy and security expertise in Q7.

Our participants are not fully representative of the global users of VPNs. Our respondents skewed older, male, and more educated than the general U.S. population; this reflects the main user population for VPNs, especially in the U.S.~\cite{usage-stats}. Our collaboration with \redacted demonstrated to us that their user base, who formed a large part of our recruitment, are avid VPN users that express the need for recommendations and advice from experts. Though we study a more-educated and possibly more tech savvy user base, the issues that we identify in our results (\eg, inadequate understanding) lead us to believe that such problems may be \textit{even more prevalent} among the general U.S. population. Therefore, we argue that our results serve as an upper bound, and our recommendations will benefit the larger, more-general user base as well.

We restricted our analysis to only people located in the U.S. While VPN users outside the U.S. have diverse and valuable perspectives, their use cases are also different. Future studies could specifically explore the perspectives of users from countries where VPNs are commonly used to circumvent censorship or other access restrictions.

We intentionally only include users of commercial VPNs, university VPNs (typically managed by the university or a third-party), and users of free, and non-commercial VPN services in this study. We do not include users of self-hosted VPN solutions or (managers and users of) workplace-specific VPNs. We leave it to future work to explore these specific subgroups of users, since they are typically more highly-skilled, and/or possess high levels of technical knowledge.
\section{Data Characterization and Validation}
\label{sec:data}
\begin{table}[t]
    \centering
    \scriptsize
    \begin{tabular}{lrr}
    \toprule
        \textbf{Demographic}              & \textbf{Respondents} & \textbf{\%} \\
        \midrule
    	Man       & 1011 & 80.75\%    \\
        Woman   &    202 & 16.13\% \\
        Prefer not to disclose & 35 & 2.8\% \\ 
        Non-Binary & 4   &    0.32\% \\
        \midrule
    	Over 65 & 741 & 59.19\% \\ 
    	56-65 & 260 & 20.77\%  \\
    	46-55 & 105 & 8.39\% \\
    	36-45 & 56 & 4.47\% \\
    	26-35 & 36 & 2.88\% \\
    	Prefer not to disclose & 34 & 2.72\% \\
    	18-25 & 20 & 1.6\% \\
    	\midrule
    	Post-grad education & 527 & 42.09\% \\
    	College degree & 508 & 40.58\% \\
    	Some college, no degree & 150 & 11.98\% \\
    	High school or eqlt & 41 & 1.20\% \\ 
    	Other & 15 & 1.2\% \\
    	Prefer not to disclose & 11 & 0.88\% \\
     	\midrule
     	High-expertise users (Knowledgeable/Expert) & 511 & 40.81\% \\
     	Moderately knowledgeable users & 631 & 50.40\% \\
     	Limited-expertise users (No or mildly knowledgeable) & 110 & 8.79\% \\
     	\midrule
     	\midrule
     	Total & 1252 & \\
    	\bottomrule
    \end{tabular}
    \caption{Demographics of the (n=1252) survey respondents.$\clubsuit$}
    \label{tab:demographics}
\end{table}

\paragraph{Survey Responses.} In total, we collected the user survey responses for six months, from March 1 to September 1, 2021. We had a total of 1,514 valid, completed responses out of which 1,374 (90.8\%) indicated they are in the U.S.. The second-highest country (China) had 23 participants, and 20 countries had only 1 participant each. We decided to focus on the U.S.-based participants (and VPN providers popular in the U.S.) as there is not enough sample to draw meaningful conclusions about other countries.

\paragraph{Quality Checks.} We have three questions, one quality- and two attention-checks, to ensure high-quality responses. Among the 1,374 U.S.-based participants that finished the survey, 1,264 or 92\% passed our generic quality check. Next, we filter out users that failed both of our attention checks (Q11 and Q21). Furthermore, we review open-ended responses from the 259 participants that failed at most one attention check, as done in~\cite{mayer2021now}, and find that over 95.8\% of these users had insightful responses. Hence, we consider \textbf{1,252 users} that passed at least one attention check for the rest of the analysis.

\paragraph{Participant Demographics.} Ours is the largest survey of VPN users to date, and we report on the 1,252 valid, high-quality responses. Shortly after launching the survey, we served on the panel of a VPN workshop organized by \redacted with over 1,500 enthusiastic users in attendance, and sent out our study recruitment message to them. We believe that our various recruitment methods ensure that we study users who are highly motivated about commercial VPNs and actively use them. Our participants skewed older, male, and highly educated. However, due to the high number of responses we obtained, we are still able to make significant conclusions from the data. Though our participants do not represent \textbf{all} VPN users, our results (\S\ref{sec:results}) indicate concerning issues even amongst the more educated, more tech savvy users, implying that our recommendations likely will benefit the more general VPN user population. The demographics are described in Table~\ref{tab:demographics}.

\begin{figure}[t] 
 \centering
 \includegraphics[width=0.95\columnwidth]{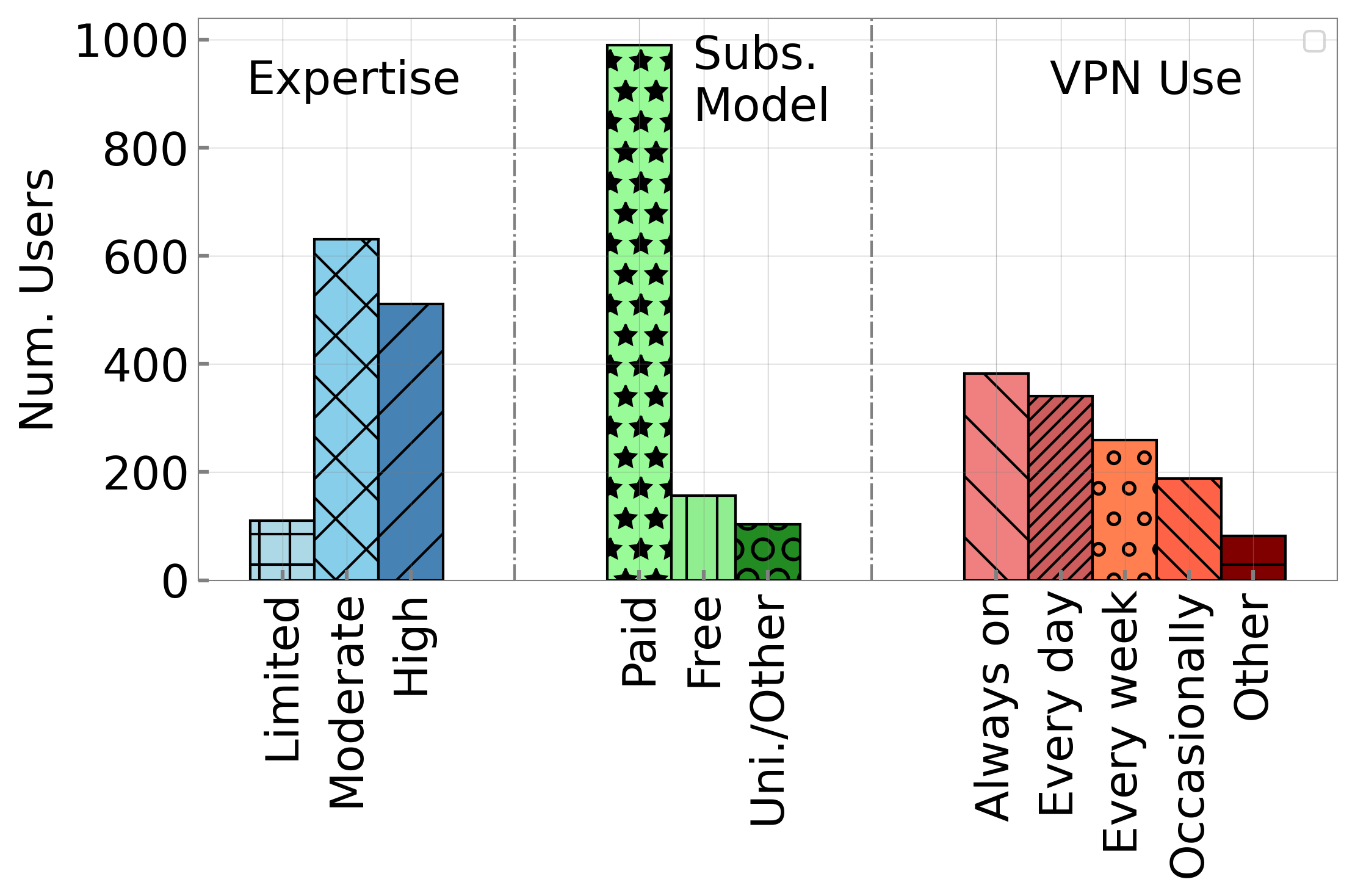}
 \caption{Overall statistics of the users' security and privacy expertise, the general VPN subscription type they use, and VPN usage.$\clubsuit$}
\label{fig:basic-stat}
\end{figure}

\paragraph{Cross-validating Self-reported Expertise.} We report our results for different sub-groups of users based on their self-reported expertise, and type of VPN subscription they generally use, shown in Figure~\ref{fig:basic-stat}. We bucket participants based on their reported expertise in security and privacy: \expert users (knowledgeable, expert), \moderate users, and \notomild users (no, mild). In order to mitigate self-reporting biases, we follow all the recommended survey design methodology best practices by including descriptive explanations for each expertise level. We craft these explanations using our expertise and incorporating feedback from user survey practitioners. We use the terms ``security'' and ``privacy'' in these descriptions to allow users to use their own judgements, and we use our threat- and mental model questions later to have the user expound on their definitions. Furthermore, we analyze the open-ended text box responses to cross-validate users' expertise: we find that \expert users provided insightful details to add to their mental models (presented in Appendix \ref{app:exp-mental}) and \notomild users were more likely to admit they do not know what protection the VPN offers them (\S\ref{subsec:mental}).
\section{Results from the User Survey}
\label{sec:results}

Security and privacy advocates, and technologists need a deeper understanding of the VPN ecosystem, and the misalignment of understanding between the stakeholders (VPN users and providers) can be exploited by bad actors to further deepen problems in the VPN ecosystem. To investigate and illuminate such issues, in this study, we conduct quantitative and qualitative studies of VPN users and VPN providers. Based on the responses from our survey and interviews, we answer the following research questions:

\subsection{RQ1: Motivations}
\label{subsec:motivation}

First, we explore the reasons for which users use VPNs and allow them to choose multiple reasons. We provide them various options that we then distill into different categories.

\textbf{Security and privacy are the main reasons why users use a VPN.} We find that protection from threats, which we consider a \textit{security motive} (82.1\%, 1,027 of 1,252) and making public networks safer to use, which we term \textit{privacy motive} (58.4\%, 731) are the biggest reasons why users use VPNs. On the other hand, censorship circumvention (8.8\%, 110) and file sharing such as torrenting (12.1\%, 151) are among the least popular reasons. Our results are in contrast with~\cite{dutkowska2020practicing} which finds university students prefer access to content (institutional, media streaming) over privacy, possibly due to the different priorities of the user populations. The overwhelming number of users that use VPNs for protection from perceived threats indicates the successful marketing of VPNs as a panacea for all security and privacy issues in the Internet.

Furthermore, 118 users also write-in additional reasons why they use VPNs (Appendix~\ref{app:reasons}). Users mention \textit{privacy} (60.2\%, 71 of 118; from ISP, tracking, surveillance, ad targeting) , \textit{security} (12.71\%, 15), being \textit{offered the service} (10.1\%, 12; by a company, with a purchase), \textit{during travel} (7.6\%, 9), and \textit{anonymity} (2.5\%, 3) as the main reasons for use.

Since finding a suitable VPN is not a trivial task, we ask users whether they had difficulty in selecting a VPN provider. Although the responses are almost evenly spread over the difficulty scale, we find differences between users with varying security and privacy expertise shown by a $\chi^{2}$-test ($p$ = 0.004206, with N=1251). As mentioned in~\ref{subsec:user-survey-methods}, we perform pairwise z-tests ($\alpha$=0.05) with FDR-BH correction to find how different user groups relate to each other.

\textbf{\Expert users less likely to find VPN discovery very difficult, more likely to find it somewhat easy.} We find that only 3.7\% (19 of 511) of \expert users find the discovery process very difficult which is significantly less than the 7\% (44 of 631) of the moderate- and 11.9\% (13 of 109) of the \notomild users who find it so. \Expert users are significantly more likely to find the process somewhat easy (21.1\%, 108 of 511, compared to 11\% of the \notomild users). 

Furthermore, we find significant difference between users that use different subscription types (free, paid/premium, other) also shown by a $\chi^{2}$-test ($p$ = 0.000005, with N=1249). Understandably, university and ``other'' VPN users (most use a VPN provided as part of a software suite) are significantly more likely to say the process was somewhat or very easy (58.8\%, 60 of 102) compared to 33.7\% (334 of 990) of paid VPN users and 28\% (44 of 157) of free VPN users. A portion of both the free VPN (40.8\%, 64 of 157) and paid VPN users (34\%, 337 of 990) find the process at least somewhat difficult. All of these findings are detailed in Table~\ref{tab:difficult} in the Appendix.

\subsection{RQ2: Needs and Considerations}
\label{subsec:needs}

To understand the needs that different users have, we ask them choose and rank criteria that they look for in a VPN. We ask the users to select the criteria they require in a VPN, and/or prefer to see in a VPN and then ask them to rank those criteria, from most important to least.
\begin{figure}[t] 
 \centering
 \includegraphics[width=0.99\columnwidth]{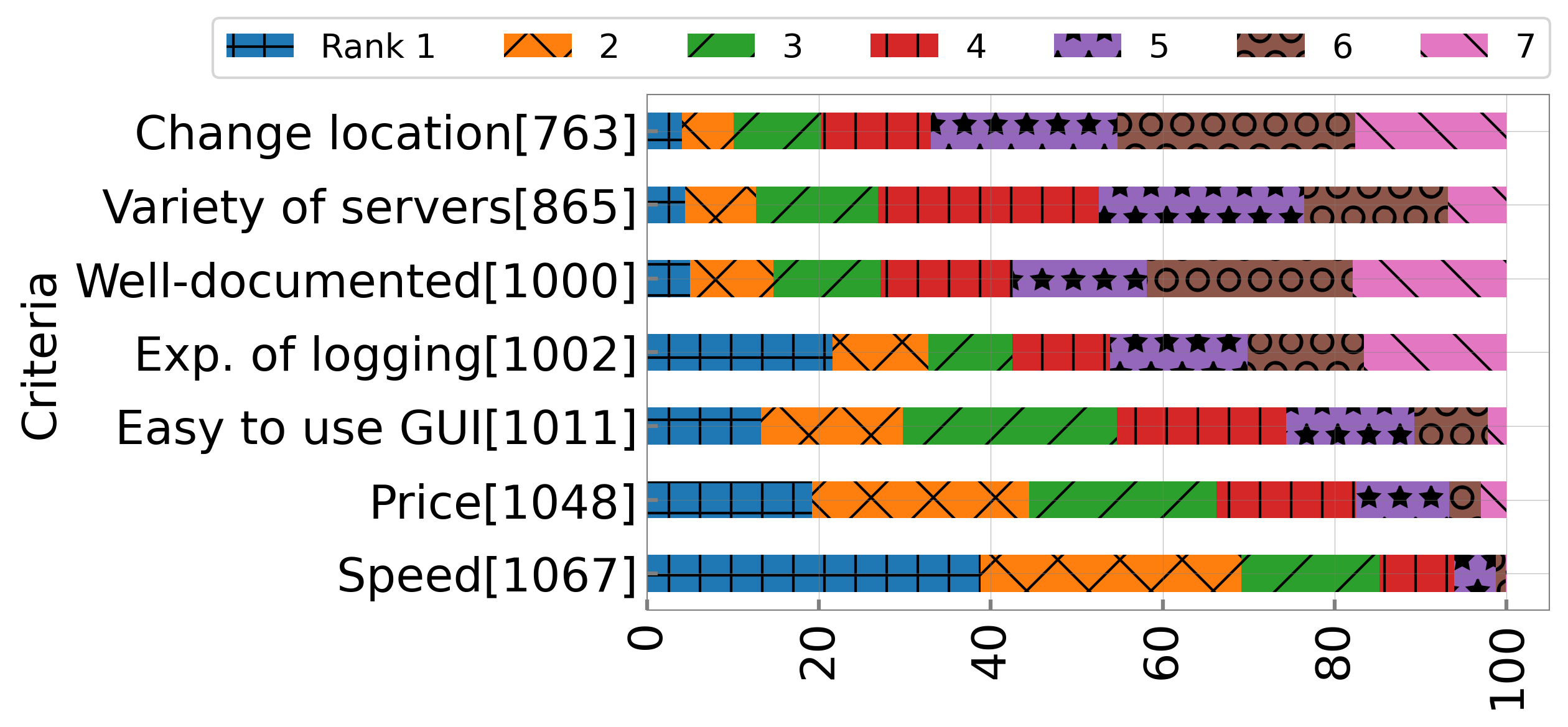}
 \caption{Importance levels users attach with criteria they look for in a VPN, presented along with number of users who chose it. Ranked from 1-most important to 7-least.$\clubsuit$}
\label{fig:rank}
\end{figure}

\textbf{Speed, price, and an easy to use app are among the top three requirements in a VPN.}
We see that speed (72.6\%, 909 of 1,252), price (55.4\%, 694), and easy to understand app/GUI (44.1\%, 553) are consistently among the top three requirements for VPN users, and over 216 users (17.3\%) ranked clear explanation of logging and data practices as their number one, as shown in Figure~\ref{fig:rank}. On the other hand, variety or number of servers (18.8\%, 235 of 1,252), and using a VPN to change location for media sites such as Netflix (12.4\%, 155) are among the lowest ranked requirements. We also find that logging data practices, which have received relatively little study, are ranked more highly than criteria like changing location for content or number of VPN servers, which have received more attention in the literature~\cite{weinberg-imc18}. We highlight that understanding real-world user requirements can help shape future research focus.

\textbf{Price is a big criteria for limited-to-moderate expertise users.} Interestingly, users of all expertise rank speed equally highly as a top three criteria (no significant differences, $p$=0.348, N=1067). But limited-to-moderate expertise users are significantly more likely to rank \textit{price} higher ($\chi^{2}$-test, $p$=0.000150, N=1048); 71.1\% (436 of 613) of these users rank it in their top three, compared to 59.3\% (258 of 435) of \expert users. This means that prices, discounts, and marketing around these factors is bound to have a vast effect on these users, similar to the study on UK and Japan users~\cite{sombatruang2020attributes}. As we will demonstrate in \S\ref{sec:providers-results}, malicious marketing around pricing is common and dark patterns are often used to ensure customer lock-in.

On the other hand,\textbf{ \expert users rank clear explanation of logging significantly higher} (53.4\%, 237 of 444 who chose it put it the top three) than all other users (33.8\%, 164 of 485 moderate- and 34.2\%, 25 of 73 \notomild users) as shown by a $\chi^{2}$-test ($p \ll$0.0001, N=1002). Also, we find that significantly more \expert users value an easy to understand GUI lower (only 38.6\%, 158 of 409 \expert users chose and rank it in their top three) compared to 64.4\% (334 of 519) of the moderate- and 73.5\% (61 of 83) of the \notomild users, shown by a $\chi^{2}$-test ($p \ll$0.0001, N=1011). This indicates that \expert users may be more confident in their ability to use a VPN application, and place higher value on the clarity of communication about the VPN service and the provider's data practices.
\begin{figure}[t] 
 \centering
 \includegraphics[width=0.90\columnwidth]{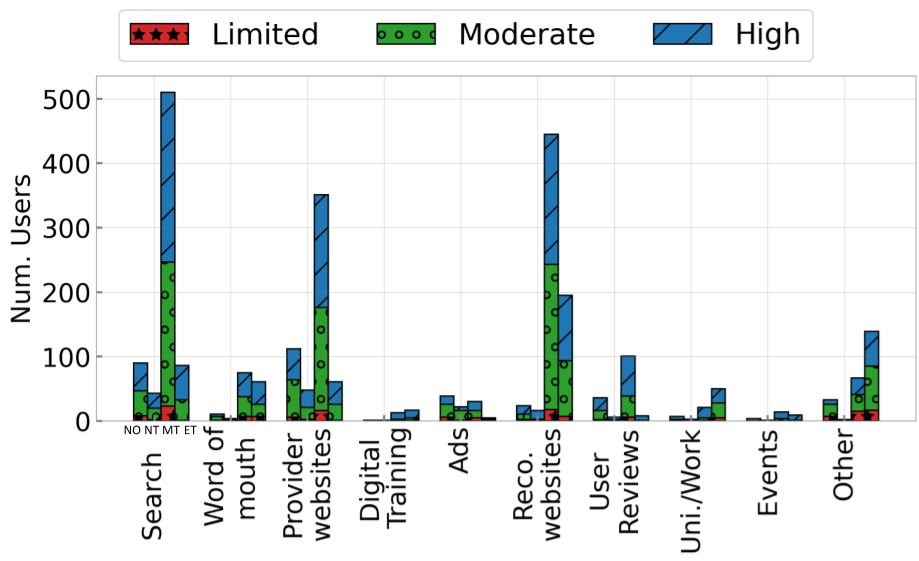}
 \caption{Trustworthiness of each resource as rated by users with different security and privacy expertise. Bars are No Opinion, Not-, Moderately- and Extremely-Trustworthy.$\clubsuit$}
\label{fig:trustworthy}
\end{figure}

\textbf{Users rely on search and recommendation sites rather than word of mouth to choose a VPN.} Given these different needs and criteria, we explore what resources users use to discover and choose the right VPN for them. Users report that actively researching on the Internet (61.1\%, 765 of 1,252), using recommendation websites (56.5\%, 708), and reading the VPN providers' websites (48.1\%, 602) are the top three ways they use to find a VPN for their needs. Users lean on these search engines and recommendation websites, rather than traditional methods like word of mouth from friends and family (5.7\%, 167), or digital training workshops (1.19\%, 35). This highlights the perils of an unregulated advertising and marketing ecosystem around VPNs, as we expound in \S\ref{subsec:stakeholders}.

\textbf{Users rate recommendation websites as trustworthy sources.}
Interestingly, among the top three resources they use, more users rate recommendation websites as trustworthy compared to the other two; 93.9\% (665 of 708) of them rate them moderately to extremely trustworthy. Figure~\ref{fig:trustworthy} illustrates how users rate the trustworthiness of each of the resources. Notably, a high proportion of users whose work or school provides their VPN service rank it extremely trustworthy (61.1\%, 55 of 90), highlighting that these users expect work/university VPNs to be of a high-quality.

Interestingly, 281 users use the ``other'' option to write-in other resources they may have used. From our qualitative coding of these responses, we notice that the VPN being offered as part of a software/security suite is the most common response (36.3\%, 102 of 281). Other responses include: trusted service provider recommendations (9.6\%, 27), and prior experience (5.3\%, 15). Appendix~\ref{app:resources} contains all the codes.

\subsection{RQ3: Emotional connection and Threat model}
\label{subsec:emotional-threat}

To understand if users attach emotional considerations such as a feeling of safety with using a VPN, we first ask them their perception of safety when browsing without a VPN and then, with a VPN. We find that there are significant differences between users that use different VPN subscription types (paid, free, and university and other) and their perception of safety without a VPN, ($\chi^{2}$-test, $p$ = 0.0001, N=1250). We explore differences between users with varying expertise levels in Appendix~\ref{app:emotional-expertise}.
\begin{table}[t]
    \scriptsize
    \centering
\begin{tabular}{rll|ll}
    \toprule
         \textbf{Population} & \multicolumn{2}{c}{Safety without VPN} & \multicolumn{2}{c}{Safety with VPN} \\
       \textbf{Subscription}  & VS/SS/NO/\textbf{SU/VU} & \textbf{U\%} & \textbf{VS/SS}/NO/SU/VU & \textbf{S\%} \\
        \midrule
        Paid/premium & 27/243/73/\hl{\textbf{(491/156)}} & \hl{\textbf{65.4}}\textupstep & \hl{\textbf{(350/538)}}/44/38/20 & \hl{\textbf{89.7}}\textupstep\\
        Free & 9/37/20/(78/13) & 58.0 & (22/97)/\hl{\textbf{24}}/11/3 & 75.8\\
        Uni.\&Write-in & 10/36/12/(37/8) & 43.7 & (35/42)/18/7/0 & 75.5\\
    	\bottomrule 
    \end{tabular}
    \caption{Number and \% of users with different subscription types and their feeling of safety without and with a VPN (from VS-Very Safe to VU-Very Unsafe).  \textupstep indicates more likely than the other subgroups for that column.$\clubsuit$}
    \label{tab:safe-paid}
\end{table}
\textbf{Users indicate they feel unsafe without a VPN, especially those who use paid/premium VPNs.} Overall, users indicate that they feel unsafe (62.6\%, 784 of 1,252) browsing the Internet without a VPN. Interestingly, we find that paid/premium VPN users are significantly more likely to feel at least somewhat unsafe when browsing without their VPN (65.4\%, 647 of 990) as compared to users that use university and other VPNs (43.7\%, 45 of 103).

\textbf{Paid VPN users are more likely to feel safe with their VPN, while free VPN users likely to indicate no opinion.} Subsequently, there are also significant differences between users with different subscription types and their perception of safety with a VPN, ($p \ll$ 0.001, N=1250). While large sections of all populations feel somewhat or very safe using a VPN (86.7\%, 1,086 of 1,252), we find that paid/premium users are significantly more likely to indicate they felt safe when using their VPN (89.7\%, 888 of 990), compared to free VPN users (75.8\%, 119 of 157) and university/other users (75.5\%, 77 of 102), who are significantly less likely. Free VPN users are significantly more likely to indicate no opinion about security (15.3\%, 24 of 157) as compared to the 4.4\% of paid users alone (44 of 990), shown in Table~\ref{tab:safe-paid}. Overall, we find that a large number of users attach emotional considerations such as safety with VPN use, and hence are likely to continue using VPNs, according to prior work studying retention~\cite{namara2020emotional}.

\begin{figure}[t] 
 \centering
 \includegraphics[width=0.90\columnwidth]{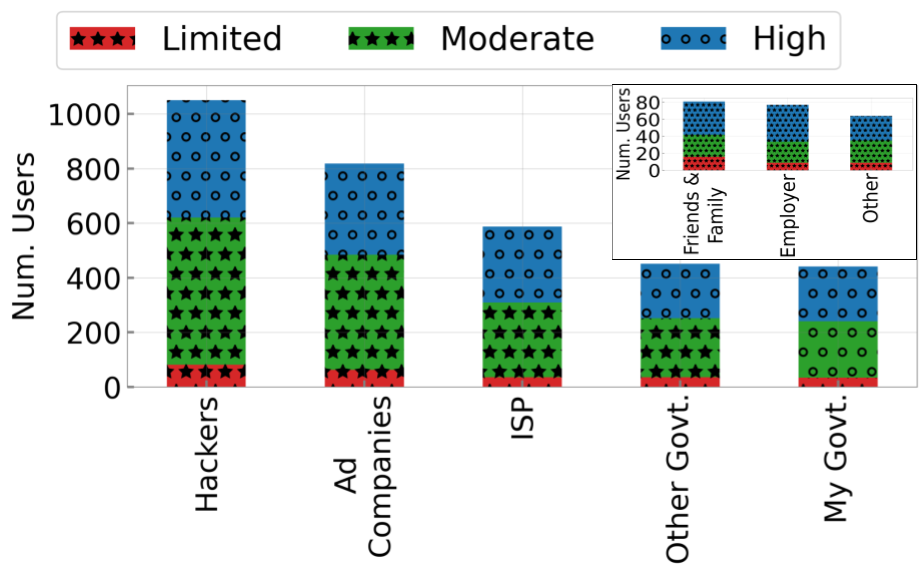}
 \caption{Entities from whom users with different security and privacy expertise want to protect their online activity.$\clubsuit$}
\label{fig:secure-reasons-high}
\end{figure}
\textbf{A majority of users use VPNs to protect and secure their online activities.} To understand users' threat models when it comes to using a VPN, we first ascertain whether users use a VPN to secure their online activities, and if yes, who they want to protect it from. Notably, 91.5\% (1145 of 1,252) of users indicate they use VPNs for securing or protecting their online activity. When exploring who users aim to protect themselves from, we find that hackers/eavesdroppers on open WiFi networks (83.9\%, 1,051 of 1,252), advertising companies (65.4\%, 819), and internet service providers (ISP) (46.9\%, 587) are the top three responses. Notably, only $\approx$30\% of users are concerned about the U.S. government or other governments. This is intriguing because post Snowden's surveillance revelations in 2014, more users moved towards privacy tools such as VPNs and anonymity tools such as Tor~\cite{smith2016snowden}. Our results indicate a shift in user's attitudes, and show a growing concern towards corporate and advertisement surveillance. This shift could be due to the influence of the marketing around VPNs and the security advice to which users are exposed. Prior work also shows YouTubers often cite ``the media'' and ``hackers'' as common adversaries~\cite{akgulinvestigating}. Figure~\ref{fig:secure-reasons-high} shows the number of users for each of these options.

\textbf{\Expert users more likely to list their ISP in their threat model.} We test each option independently to see if there are significant differences between users with varying expertise. We find that significantly more \expert users indicate their ISP as one of the reasons (54.4\%, 278 of 511), as compared to other users (43.3\%, 273 of 631 moderate-, and 32.7\%, 36 of 110 \notomild users) ($p \ll$ 0.0001, N=1252). While no significant difference was found between users selecting advertising companies ($\chi^{2}$, $p$=0.157, N=1252), significantly less proportion of \notomild users indicate that hackers and eavesdroppers are a concern (73.6\%, 81 of 110) as compared to 85.6\% (540 of 631) of the moderate- and 84.1\% (430 of 511) of the \expert users, as confirmed by a $\chi^{2}$-test ($p$=0.00695, N=1252).

\subsection{RQ4: Mental Model}
\label{subsec:mental}
To evaluate users' mental model of VPNs, we ask them a scenario question which aims to elicit their understanding of what protection a VPN actually provides. In the given scenario (Q19), the user concluding that their ISP learns what websites they visit while connected to a VPN indicates a flawed mental model. 

\textbf{Almost 40\% of users have a flawed mental model.} We find that a high portion of users (39.9\%, 500) have a flawed mental model and believe their ISP can see the websites they visit over the VPN. Worryingly, we see no significant difference between users of different expertise based on the $\chi^{2}$-test ($p$=0.0927, N=1252). Our results are concordant with previous work which find that users, even experts, have misconceptions about the protections certain tools offer~\cite{story2021awareness, binkhorst2022security}. We initially also considered the 135 users who answered ``Nobody [can see what website I visit]'' as having a flawed mental model. But we instead opt for a conservative approach and did not include them because four users clarified their response using the textbox accompanying this question. They state that since their VPN says no logging, tracking, or sharing, \textit{ideally} nobody should know what website they visited.

\textbf{\Notomild users are more likely to have an unclear mental model, while \expert users more likely to add insightful details.} We find significant difference between users that chose ``I don't know'' to this question, based on a $\chi^{2}$-test ($p \ll$ 0.00001, N=1252). We find that users with \notomild are more likely to choose ``I don't know'' (30.9\%, 34 of 110 users) compared to 16.3\% (103 of 631) of the moderate- and 5.5\% (28 of 511) of \expert users. \Expert users are significantly more likely to use the ``other'' option and write-in their answer (14.9\%, 76 of 511) as compared to 8.9\% moderate- and 1.8\% \notomild users ($p$= 0.00003, N=1252). Analyzing these write-in responses, we find that \expert users add insightful details such as \textit{DNS providers} knowing what websites user visits, and \textit{site owner} learning about the user using logins or cookies. They identify other threat actors such as the site's partners, search engine used to navigate to the site, government agencies, and browser fingerprinters; all codes are present in Appendix~\ref{app:exp-mental}. 

To understand if VPN users have a good idea about the data VPNs can collect about them, we present many options and ask users to indicate the various kinds of data they think a VPN provider collects about them. During the analysis, we bucket these options into: \textit{typical}, \textit{dangerous-unreasonable}, \textit{miscellany}, \textit{not sure}, and \textit{custom input}. While the last two are self-explanatory, ``typical'' includes demographics and account holder information, VPN servers connected to, timestamps at when VPN is in use, and device type. We consider them typical since the data is readily available to a VPN provider. The ``dangerous-unreasonable'' bucket includes: private messages, audio/video recordings, and keystrokes from device, all of which are not usually collected by a VPN provider, unless they are operating a malicious service, while ``miscellany'' includes website visited, geolocation, and interests for ads. While a reasonable provider would not collect this type of data, it is possible that some VPN providers do collect them.

\begin{table}[t!]
    \scriptsize
    \centering
\begin{tabular}{rrrrrr}
    \toprule
       \textbf{Expertise}  & NotSure & \textbf{NS\%} & Typ/Dang/Misc/O. & \textbf{Typ.\%} & \textbf{Dang.\%} \\
       \midrule
       High & 132 & 25.83  & 326/35/217/58 & 86.02 & 9.23 \\
       Moderate & 304  & 48.18 & 292/44/220/32 & 89.30 & 13.46\\
       Limited & 68 & 61.82 & 35/18/36/3 & 83.33 & 42.86 \\
    	\bottomrule 
    \end{tabular}
    \caption{Number and \% of users who indicate the types of data they think VPN providers collect. Users can choose multiple options, and we exclude users who chose ``not sure'' (NS) from the other counts.$\clubsuit$}
    \label{tab:data-collect}
\end{table}
\textbf{At least 40\% users indicate they are unsure what data is collected, and $\approx$13\% of the remaining users think unreasonable kinds of data are collected by VPNs.} We find that 40.3\% (504 of 1,252) of users indicate they are not sure what data is collected, limited-(61.8\%, 68 of 110) and \moderate users (48.2\%, 304 of 631) are significantly more likely to indicate uncertainty as compared to 25.8\% (132 of 511) of the \expert users ($\chi^{2}$, $p \ll$0.0001, N=1252). We exclude these users from the analysis and from the remaining 748 users, we see that in general users believe typical data (87.3\%, 653) is collected by VPN providers. However, 13\% (97 of 748) of users think VPNs collect dangerous-unreasonable data. The fact that users of all expertise levels have this belief, reiterates the need for better, more effective user education. Table~\ref{tab:data-collect} summarizes these results.

Finally, we explore the reasons why users think such data is being collected by VPN providers. A majority of respondents (79.2\%, 992) believe the main reason is for internal analytics and quality of service reasons. Interestingly, significantly more \notomild users believe that the data is being collected for advertising (36.4\%, 40 of 110), as compared to 20.4\% of moderate- and 16.4\% \expert users ($\chi^{2}$,$p$=0.000014, N=1252). A significantly high portion of \notomild users also believe data is used for user tracking (36.4\%, 40, $p$=0.019), and selling to third parties (33.6\%, 37, $p\ll$0.0001), highlighting that \notomild users believe VPNs use data collected about users for monetary benefit.

\subsection{RQ5: Perception and Trust}
\label{subsec:perception}
\begin{figure}[t] 
 \centering
 \includegraphics[width=0.95\columnwidth]{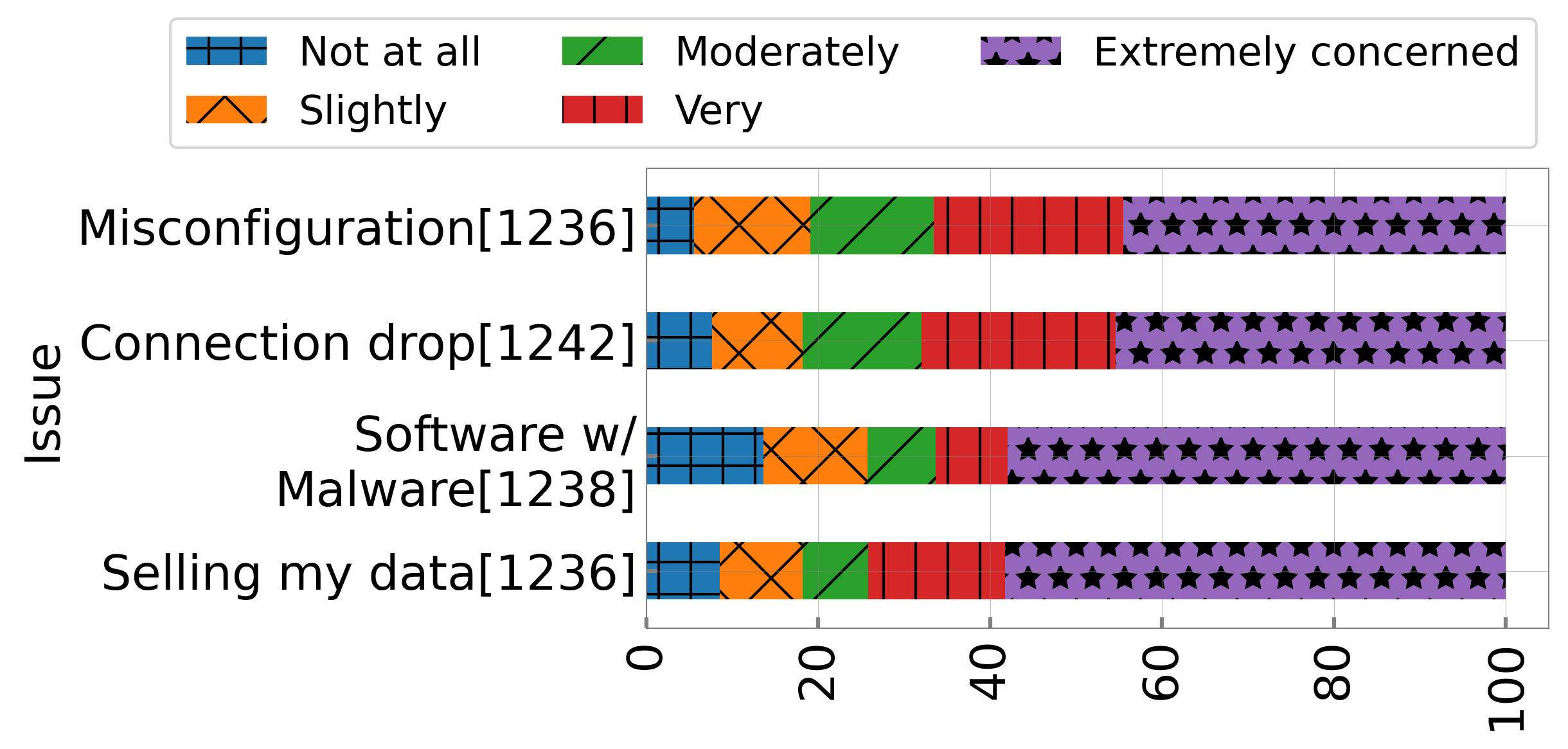}
 \caption{Users indicate their concern levels towards VPN-related issues, with the number of users who answered each.$\clubsuit$}
\label{fig:concerns}
\end{figure}
In order to understand users' perception of the VPN ecosystem and its issues, we ask users to rate their concern levels towards VPN related issues. We find that users are very or extremely concerned about VPN providers selling their data (73.2\%, 917 of 1,252), and the VPN software containing malware (65.6\%, 821). Users also express higher degrees of concern towards more technical issues such as VPN software failing without warning (67.4\%, 884), and misconfigured VPN services (65.7\%, 823), illustrated in Figure~\ref{fig:concerns}. We find no statistically significant differences between users of different expertise or subscription types for these options ($\chi^{2}$, $p$>>0.05).

\begin{figure}[t] 
 \centering
 \includegraphics[width=0.99\columnwidth]{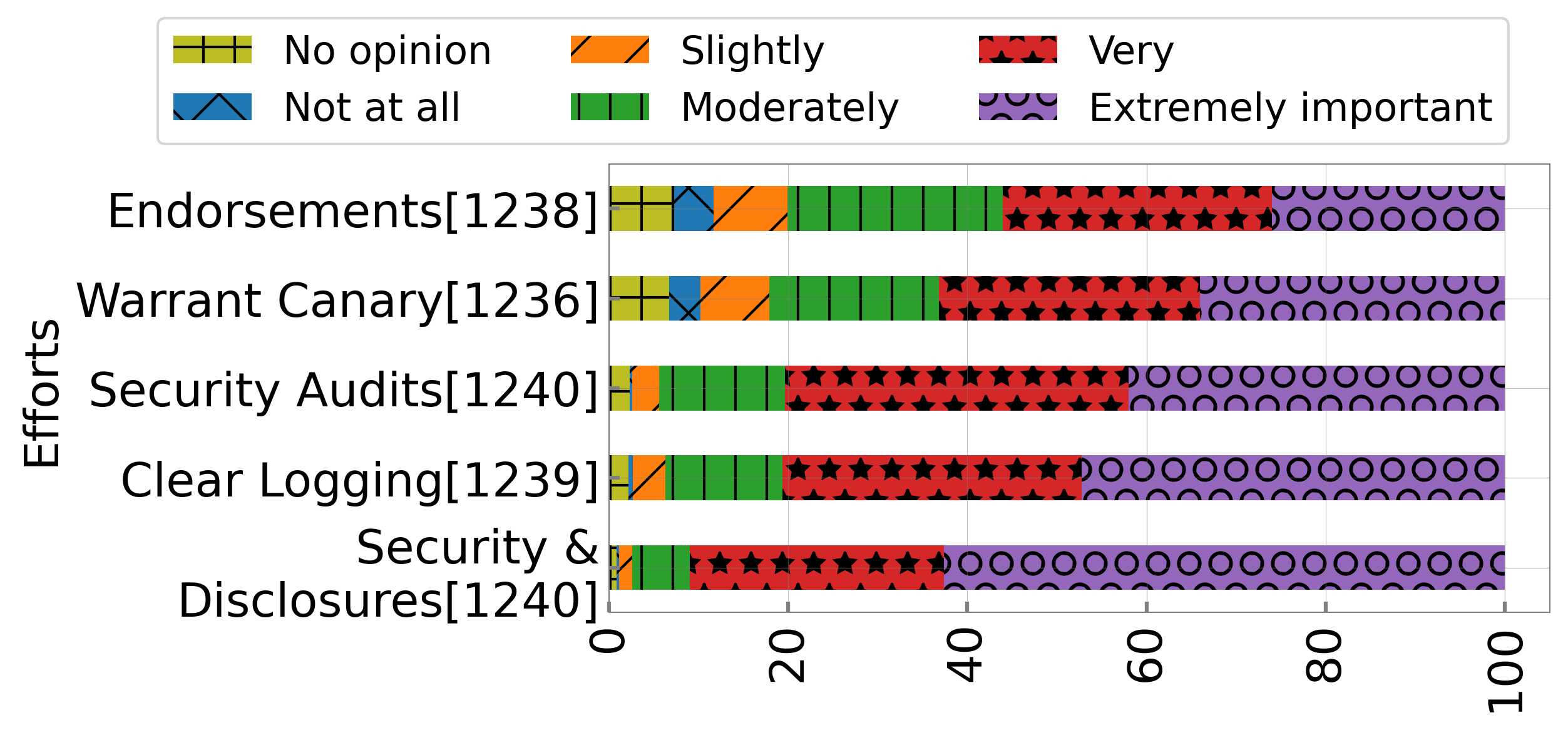}
 \caption{Users indicate the importance of trust-increasing efforts by VPNs, with number of users who answered each.$\clubsuit$}
\label{fig:important}
\end{figure}

Finally, we ask users what level of importance they associate with efforts that VPN providers undertake to earn and increase trust from the user base. We find that users consistently rate security protocols and disclosure of breaches (62\%, 776 of 1,252) as an extremely important effort, followed by having a clear logging policy (46.7\%, 585), and independent security audits (41.6\%, 521), as shown in Figure~\ref{fig:important}. While there may be other efforts that we do not list, we hope that VPN providers and researchers use these insights gleaned from the users' perspectives to inform their future efforts and campaigns to secure and foster user trust.
\section{Perspectives of the VPN Providers}
\label{sec:providers-results}

In this section, we present exploratory results from our VPN provider interviews and summarize the key issues and themes, with number of providers per theme in brackets. We compare these insights with results from the user survey, and highlight the key areas where the two are misaligned.

\subsection{Key Themes}
\paragraph{Key Efforts.} We learn from providers that they focus on cross-platform security development (6/9), product simplicity (4/9), and usability (5/9) of their product. They also mention that they try to be reliable, gain trust over time (5/9), and practice transparency (5/9). We also noticed many VPN providers mentioned offering additional features, such as filtering, ad- and tracker-blocking similar to anti-virus software, indicating that VPNs are evolving beyond their normal functionality to retain users. From a mental model perspective, this could potentially be harmful as it sets an over-expectation of security and privacy, while users are already unclear about protections that standard VPNs offer them.

\paragraph{High-level Challenges.} When asked about the biggest challenges in the industry, providers explain that \textit{building trust} (6/9) is hard because there is a large number of providers and little transparency. We find that providers agree that problems generally stem from lack of trust, focusing on features and not privacy, and overestimation and overselling of service. Providers also mention that users do not understand risks (7/9), and that it is their responsibility to do better in user education and ensure honesty in their disclosures to users.

\paragraph{User Base.} When asked about their user base and whether they conduct studies to understand them, almost all providers explain that having a \textit{privacy-focused service deters user studies}, and that they try not to learn about their users (7/9). Instead, they typically depend on inbound user feedback such as in-app surveys or support tickets. We notice that commercial providers mention that they prefer privacy-centric users, and that western users are more likely to be paying customers.

\paragraph{Pricing \& Marketing.} Providers mention that development, labor and marketing are the main factors affecting pricing (5/9). Other factors include deals with server and cloud providers, organization build, technical means, and infrastructure. They mention that growing the user base is imperative as it creates economies of scale. Providers also note the existence of malicious practices around discounts that are not user-friendly (5/9), like marketing gimmicks to lock users. Multiple providers remark that it is the norm of the industry (3/9), one of whom says:
\begin{quotation}
``I think it's not good for consumers but why everyone does it, because everyone else does that.''
\end{quotation}
\noindent A majority of providers agree that marketing plays a big role; noting that the marketing costs are high, and the competition is harsh. Regarding marketing methods, many providers mention that they do ethical marketing by being involved with the user community, relying on user reviews and word of mouth.

\paragraph{VPN Review Ecosystem.} We discover that a main theme from the interviews is the issue of the VPN review ecosystem. One provider calls it a ``parasitic industry'' and a majority of providers (6/9) remark that the review ecosystem mostly runs on money, \eg paid reviews, and cost-per-action (CPA). They also explain that VPNs or their parent companies may own different review sites~\cite{reclaimthenet-kapetech}, many review sites even \textit{auction the \#1 spot}, and do reviews for money. 
Multiple providers also mention that Google search results are unreliable, and that there are few good reviewers left; one provider says:
\begin{quotation}
``You honestly cannot find even one ranking site that is honest, if you just tell people that...so that people know''
\end{quotation}

\paragraph{Dark Patterns in the Industry.} Another recurring theme was about dark patterns in the industry. Since most of these patterns are usually not readily apparent to users and researchers, we also explicitly ask a question about them. We divide the issues mentioned by various providers into:

\textit{Operational Issues (7/9):} These include VPN providers having anonymous or unknown owners, having deceptive subscription models, and tracking users on their own sites, which was also highlighted in a recent report~\cite{markup-vpntracking}. Providers also remarked on aggressive and unethical marketing such as retargeting users with VPN ads, and relying on users forgetting to cancel subscriptions. On the other hand, providers mention that VPNs get attacked as well (by other providers, bad users, and by those who abuse free VPN services).

\textit{Malicious Marketing (6/9):} Providers mention several issues, that we term as \textit{malicious marketing}, including the use of affiliate marketing, preying upon users' lack of knowledge, and overselling of service including selling anonymity even though that is not a VPN guarantee. They also foster a false sense of security around VPNs through misinformation, fearmongering, dishonest non-expert reviews, and lying to users in disclosures. One provider, on fearmongering:
\begin{quotation}
``The best ways to get people to pay for something is to scare them and to tell them that they need security''
\end{quotation}

\textit{Factors Enabling Dark Patterns (4/9):} Providers bring up several challenges that exacerbate these practices, such as the fact that the VPN ecosystem has no accountability, lacks transparency, and has few marketing and advertisement standards. Since the VPN industry is spread over multiple jurisdictions, it is hard to regulate. One provider calls it \textit{the wild west}:
\begin{quotation}
``You know we could just say literally anything...there's absolutely no oversight. There's no one to tell you, \textit{``Ah, you can't say that because that's not true.''} There's no regulation, there's no kind of governing body''
\end{quotation}

\subsection{RQ6: Alignment between VPN users and providers}
\label{subsec:stakeholders}
We highlight several key areas where VPN users and providers are misaligned in their understandings and incentives, in addition to issues that both parties agree on. By highlighting these issues, we hope that technologists, and security advocates prioritize users' challenges, and focus on key problem areas. We arrange these issues from most aligned to least. 

\paragraph{Privacy-centric Users.} We note that providers explicitly mention that they prefer and cater to privacy-centric users, which aligns with the findings from our survey where over 91\% of users mention that they use VPNs for security and/or privacy. Since providers mention they respect privacy and are unable to conduct user studies of their own, it is imperative for researchers to develop an understanding of VPN users.

\paragraph{Users' Mental Model of VPNs.} Providers say that users have flawed mental models of VPNs (6/9) and our survey concurs that $\approx$40\% of users do indeed have a flawed mental model. Providers and the security advocacy community should hence place high priority on user education. Providers mention that challenges in improving users' mental models include the lack of positive reinforcements (visual signs that a VPN is working), constant exposure to negative experiences (increased encounters of CAPTCHAs, media sites blocking VPN use), and striking a balance in technical communication. We emphasize that user-onboarding, clear communication, and responsible advertising are key drivers for change.

\paragraph{Importance of Pricing.} From our user survey, we see that pricing is among one of the highest priorities for users, especially for limited-to-moderate expertise users. However, providers on the other hand mention that certain malicious marketing gimmicks are often used---such as fearmongering, fake countdown timers, and being always on sale---to lock-in users. We fear that since pricing is key for users, malicious tactics used by certain providers may chain users to a service that may not necessarily meet security standards. We strongly urge that advocates focus on regulations to protect consumers.

\paragraph{Users' Reliance on Review Sites.} Despite most providers agreeing that the review ecosystem is not objective about the services and is instead largely motivated by money, our survey shows that users strongly rely on them and believe they are trustworthy. Though our survey studies only U.S. users, the VPN providers believe that the western population (including U.S.) are more likely to pay for their subscriptions. It is important to deter the exploitation of these users by informing them of the nature of the review ecosystem and how the reviews and rankings are made. As we highlight from the providers' interviews, a lot of the malicious marketing preys on users' misunderstandings. Hence, shedding light on these behaviors in the review ecosystem is crucial to ensure that they do not continue profiting off users via paid reviews and CPA. One provider says:
\begin{quotation}
``[Running costs have reduced] in the last 10 years, yet [VPN] prices are all the same. Why is that? Well it's because the VPN review sites are getting all the money.''
\end{quotation}

\paragraph{Users' View on Data Collection.} We find that over 40\% of users are not sure exactly what data is being collected about them by VPN providers. Of the remaining users, we find that 13\% think that VPNs collect dangerous or unreasonable kinds of data. On the other hand, multiple VPN providers say that they clearly communicate their logging practices, or that they do no logging and have audits to prove it. From our survey, we also find that having a clear logging policy is among the top important indicators for increasing trust with users. Alongside improving users' mental models of how VPNs work, this is another key issue that VPN providers can address by better informing users about their operation.

\section{Actionable Recommendations}

Traditional approaches to regulating including standardization by government bodies may not be the best solution for VPNs because the providers and VPN servers span multiple jurisdictions. Another approach can be self-regulation within the industry. However, though coalitions look good on paper, it is necessary to bring enough providers together, and ensure oversight in order to hold these coalitions accountable. One provider, on why having such an alliance is hard: 
\begin{quote}
    ``[VPN providers] don't want to be held accountable for the [mistakes] of other providers...there's not a lot of trust.''
\end{quote}
Even if providers do form coalitions, we find that they do not really hold to their own self-regulated principles. In our prior work, we also find that the lack of regulation and standardization leads to VPN providers offering varying levels of security and privacy~\cite{vpnalyzer-ndss}.

We strongly recommend that FTC and other government organizations exert oversight on VPN advertising and curb malicious tactics used by VPNs, because such aggressive and misleading ad campaigns could degrade users' mental models about VPNs. An example of successful oversight is NordVPN's ad being banned in the UK for misleading users~\cite{nordvpn-register}. In addition, we advocate for coordinated efforts from the industry, academia, and consumer protection organizations to bring attention to the flawed VPN recommendation ecosystem.

Finally, our study also shows that user education campaigns regarding VPNs and the VPN ecosystem must be prioritized. We find key areas that need the most improvement: users' mental model of what a VPN provides, what data it can collect, and the threat models for which VPNs can be most useful. Since the user population surveyed in our study is on average older and more educated, our results suggests that incomplete and flawed mental models may be even more prevalent among the general U.S. population. We urge security and privacy advocates such as the EFF and CDT, consumer protection agencies such as the FTC, and community initiatives such as IFF to devote their efforts towards VPN user education, raise awareness, and advocate for VPN industry oversight. 

\section{Discussion \& Conclusion}
\label{ref:discussion}

VPNs have quickly gained popularity as a security and privacy tool for regular Internet users. Commercial VPNs are now a multi-billion global industry with numerous VPN providers, and apps on almost every platform. In our interviews with them, multiple providers mention that setting up a VPN and offering a service is not technically difficult, especially with the existing open source solutions~\cite{openvpn,donenfeld2017wireguard}, and highlight that many VPN companies have unknown or anonymous ownership. One provider says there is a low bar to entry:
\begin{quotation}
    ``Technically it's not that hard to run a VPN...two people in a basement with a half decent power....can run a VPN.''
\end{quotation}

For users however, exposure to risk of surveillance, reports of ISPs selling data, and increasing access restrictions have all led to an increased awareness of online risks. VPNs are marketed as technological solutions to many of these issues, though not all users will be able to verify these claims. In simple terms, a user using a VPN is simply transferring trust, say from their Internet provider, onto the VPN provider. Internet service providers (ISPs) have been around for longer and have many regulations globally. However, such regulations and advocacy has not yet caught up to the VPN industry. 

In this paper, we conduct studies on VPN users and providers and present actionable recommendations on important problem areas in the VPN ecosystem. Our interviews with VPN providers helps open up communication between academia and companies developing privacy-enhancing tools, which can lead to transfer of knowledge, foster collaboration, and help develop solutions for issues in the ecosystem that ultimately impacts users. We highlight that understanding real-world user needs and requirements can help shape future research focus. We hope that by shedding light on issues such as the ones rampant in the VPN review ecosystem, we raise awareness and encourage investigation, advocacy, and regulation to improve the entire VPN ecosystem for the better.

\vspace{-0.5em}

\section{Acknowledgment}
The authors are grateful to Michelle Mazurek for her valuable input, and Karen Jaffe, Philipp Winter and Jane Im for their feedback on the survey. We thank Ben Moskowitz, Leah Fischman, Yael Grauer, and the reviewers for their feedback for their feedback. This work was made possible by the Open Technology Fund, Consumer Reports Digital Lab Fellowship, and National Science Foundation grant CNS-2141512.

{\clubpenalty=0\widowpenalty=0
\bibliographystyle{abbrv}
\bibliography{paper}}
\begin{table}[h]
    \scriptsize
    \centering
\begin{tabular}{rrrr}
    \toprule
       \textbf{Population}  & VD/SD/NE/SE/VE & \textbf{V. Difficult\%} & \textbf{S. Easy \%}\\
       \midrule
       \Expert & \hl{19}/139/158/\hlc[pastelyellow]{108}/87 & \hl{3.7\textdownstep} & \hlc[pastelyellow]{21.1\textupstep}\\
       \Moderate & 44/180/198/116/93 & 7.0 & 18.4 \\
       \Notomild & 13/22/39/12/23 & 11.9 & 11 \\
       \midrule
        \textbf{Population}  & Diff/Neither/Easy & \textbf{Difficult\%} & \textbf{Easy \%}\\
        \midrule
       Paid/Premium & 337/319/334 & 34 & 33.7 \\
       Free & 64/49/44 & 40.8 & 28.0 \\
       Other (Uni./other) & 16/26/60 & 15.7 & 58.8\textupstep \\
    	\bottomrule 
    \end{tabular}
    \caption{Number and \% of users from different user groups indicate how difficult it was decide on a VPN to use (from VD-Very Difficult to VE-Very Easy). Symbols indicate \textupstep more, and \textdownstep less likely than the other rows in the column.$\clubsuit$}
    \label{tab:difficult}
\end{table}
\appendix
\section{Appendix: Emotional connection with VPN for different user expertise (RQ3)}
\label{app:emotional-expertise}
As shown in ~\ref{subsec:emotional-threat}, in general, users indicate they feel unsafe without a VPN. We find that there are no significant differences between users with varying expertise levels and their perception of safety without VPNs, as shown by a $\chi^{2}$-test ($p$ = 0.085, N=1252). We notice that less \notomild users indicate that they feel at least somewhat safe without a VPN (only 20\%, 22 of 110 as compared to 30.3\% of the high- and 29.5\% of the \moderate users). 

While this is not a statistically significant difference, we explored the reason they do not feel safe without a VPN by analyzing their textual response immediately after this question. \Notomild users who responded (98 of 110) mainly express worry (about hacking, tracking, and more), and confusion about what VPN offers, and explain scenarios where they feel unsafe. S99 says:

\begin{quotation}
``One never knows when either the so-called good guys or the bad guys are lurking about, just waiting to pounce. In my book, I want to be safe rather than sorry[...]''
\end{quotation}

\textbf{In general, users indicate they feel safer browsing the Internet with a VPN.} In a different section of the survey, we ask them about the perception of safety while using a VPN. We find significant differences between users with varying expertise levels and their perception of safety with VPNs as well, shown by a $\chi^{2}$-test ($p$=0.003, N=1252). While large sections of all populations feel somewhat or very safe (86.7\%, 1,086) using a VPN, \notomild users are \textit{significantly less} likely to indicate they felt safe using a VPN (75.5\%, 83 of 110) compared to 88.3\% (557 of 631) of moderate- and 87.3\% (446 of 510) of \expert users), summarized in Table~\ref{tab:safe}. We also find that instead \notomild users were significantly more likely to indicate they had no opinion on safety while using a VPN (16.4\%, 18 of 110), possibly due to confusion on what a VPN provides. S1153 says, who indicated no opinion says: 
\begin{table}[t!]
    \scriptsize
    \centering
\begin{tabular}{rlr|lr}
    \toprule
         \textbf{Population} & \multicolumn{2}{c}{Safety without VPN} & \multicolumn{2}{c}{Safety with VPN} \\
       \textbf{Expertise}  & VS/SS/NO/SU/VU & \textbf{S\%} & VS/SS/NO/SU/VU & \textbf{S\%} \\
       \midrule
       High & (22/133)/48/231/77 & 30.3 & (202/244)/30/21/13 & 87.3\\
       Moderate & (19/167)/44/324/77 & 29.5 & (179/378)/38/27/9 & 88.3 \\
       Limited & \hl{\textbf{(5/17)}}/13/52/23 & \hl{\textbf{20.0}}\textdownstep & \hl{\textbf{(27/56)}}/18/8/1 & \hl{\textbf{75.5}}\textdownstep\\
    	\bottomrule 
    \end{tabular}
    \caption{Number and \% of users with different security and privacy expertise and their feeling of safety when browsing without and with a VPN (from VS-Very Safe to VU-Very Unsafe). Symbols indicate \textupstep more, and \textdownstep less likely than the other rows in the column. Highlighted values indicate that they contribute to the relevant percentage.$\clubsuit$}
    \label{tab:safe}
\end{table}

\begin{quotation}
``I feel both somewhat unsafe and somewhat safe''
\end{quotation}

\section{Codes from qualitative survey responses}
\label{app:codes}
\subsection{Reasons for use}
\label{app:reasons}
\textbf{Privacy from ISP} (22), \textbf{Privacy:} Privacy (17), from tracking (10), from tracking and ads targeting (5), surveillance (3), securing browsing history (3), hiding location (2), from ads (2), selling my data (2), hacking (2), from attribution (1), banking (1), during searching (1), ISP and large companies (1), \textbf{Security:} during banking (4), hackers (4), as a principle (2), paranoia (1), confidential/sensitive data (2), OpSec (1), hackers/surveillance and bad actors (1), protection (1),  \textbf{Offered the service:} by Norton (4), free with other service (3), with router (1), by ISP (1), for low price (1), with device (1), from employer (1), \textbf{While travelling: } surveillance countries (2), protection from local actors (2), censoring countries (2), in general (2), don't trust hotels (1),\textbf{Anonymity} (3),  \textbf{Access geo-restricted content} (2), \textbf{Work with tech} (1), \textbf{Safeguard device} (1), \textbf{No-log VPN} (1), \textbf{for IPTV} (1), \textbf{For work/uni} (1), \textbf{Browsing from different locations} (1)

\subsection{Other Resources Used}
\label{app:resources}
\textbf{Part of Software/Security Suite} (102), \textbf{Trusted service provider} (27), \textbf{Prior Experience} (15), \textbf{Reviews:} \redacted (13), Offered the service for free (13), Introduced as part of my job (8), thatoneprivacyguy (7), \textbf{Own testing} (7), Trying the trial option (7), \textbf{Word of Mouth:} from technical staff (6), Recommended by service (2), Computer Clubs (2), Colleague (2), University (1), Indiegogo (1), Meetings (1), Geek Squad (1), Friend/Family (2), Computer services company (1)
\textbf{Trust:} the Mac App Store (5), the Google play store (1), Tech YouTubers (1), Leo Laporte (1), privacytools.io (1), Bloggers, Apple News+ (1), Expert reviewer (1), \textbf{No choice in VPN provided} (4), \textbf{Company Announcements} (5), 
\textbf{Reviews:} Reviewer Kim Komando (3), Specific to MacOS (3), PCMag (3), News Articles (3), Recommendation Sites (2), Trusted sources (2), NYT (2), Local tech advisor (2), testmy.net (1), ZDNet (1), Recommendation on YouTube (1), Print Magazines (1)
\textbf{Advertising:} Ads on trusted podcast (3), Promos (3), in specific site (1)
\textbf{No Research} (3), \textbf{Provider's website} (1)

\subsection{Feeling of Safety without VPN: \Notomild users}
\label{app:safety-wo-limited}
\textbf{Worry:} hacking (13), tracking (6), exposed personal details including IP and location (10), unsafe in today's world (4), bad actors (3), dark web/net (2), ISP access data (1), open to exploitation (1), malware (1), less protected (1), happened to others (1), breaches (1), ID theft (1), prevention (1), 
fear threats and financial data (1), \textbf{Confusion:} don't understand (10), what does ISP do with data (1), service stopped working (2), \textbf{Safety:} no prior issues (5), I'm careful (3), I have anti-virus (2), using VPN makes me safer (2), use trusted provider (3), my device is safe (1), I am trusting (1), added protection (1), \textbf{Scenario:} only unsafe in public networks (4), I use it if I have it (1), no reason (1), HTTPS isn't always available (1), \textbf{No worry: } I feel okay (2), \textbf{Understanding:} with research (1),  its supposed to hide me (1), anyone can see my traffic (1), \textbf{Specific needs} (1), \textbf{Needs:} trade-offs (1), harder for hackers (1), make me safer (1).

\subsection{High-expertise users response to mental model}
\label{app:exp-mental}
\textbf{DNS Provider} (7), \textbf{Site:} if entered personal info (6), has access to cookies (4), might know (5), if insecure protocol (1), \textbf{VPN Provider:} logging (5), depending on service (4), surely knows (3), alone cannot hide you (2), audit trail (1), \textbf{ISP knows if DNS leaked} (4), \textbf{Other actors:} example site's partners (3), large companies like Google/Facebook (3), Browser (2), search engine (2), ad networks (2), IDSs (1), badly implemented tech (1), \textbf{Threat actors:} tracking (3), government agencies (2), third party cookies (2), browser fingerprinters (2), hackers (1), \textbf{Idk:} nobody if no logging (2), any hop in between VPN and site (1).

\section{User Survey}
\label{app:user-survey}
We first display the survey landing page containing the consent form and introduce the survey. Then, we display the following questions:
\paragraph{Demographics \& General Questions about Internet Usage}\mbox{}\\
\textit{In this section, we ask a few general questions to collect demographic information, Internet habits, and self-reported security and privacy knowledge rating.}\\\\
\noindent Q1. Gender*\\
\emptycirc Woman \emptycirc Man \emptycirc Non-binary \emptycirc Prefer not to disclose \emptycirc Prefer to self-describe\\\\
Q2. Age Range*\\
\emptycirc 18--25 \emptycirc 26--35 \emptycirc 36--45 \emptycirc 46--55 \emptycirc 56--65 \emptycirc >65 \emptycirc Prefer not to disclose\\\\
Q3. Country of Residence*\\
\textit{Dropdown list of countries}\\\\
Q4. Highest Level of Education*\\
\emptycirc Some education, but no high school diploma or equivalent \emptycirc High school degree or equivalent (e.g. GED) \emptycirc Some college, no degree \emptycirc College or university degree (for example a bachelor’s or associate's degree) \emptycirc Post-graduate education (for example a master’s or a doctorate degree) \emptycirc Other (Please elaborate below) \emptycirc Prefer not to disclose\\\\
Q5.  Is your field of study connected to computer science and/or technology?*\\
\emptycirc Yes \emptycirc No \emptycirc Prefer not to disclose\\\\
Q6. On a general day, how many devices (e.g. mobile phone, laptop, tablets) do you use to browse the internet?*\\
\emptycirc 1 \emptycirc 2 \emptycirc 3 \emptycirc 4 \emptycirc 5 \emptycirc 6 \emptycirc 7 \emptycirc 8 \emptycirc 9 \emptycirc 10+\\\\
Q7. How would you rate your knowledge about privacy \& security on the Internet?*\\
\emptycirc 1 - No knowledge - I do not have any knowledge of privacy or security concerns pertaining to the Internet
\emptycirc 2 - Mildly knowledgeable - I’ve heard of things such as VPNs, proxies, or Tor, but don’t really understand how they work and/or have limited experience using them; I cannot describe specific dangers to privacy or security online but I know they exist.
\emptycirc 3 - Moderately knowledgeable - I can name some of the dangers to privacy online and am wary of them in general; I know about privacy tools and have tried using some of them.
\emptycirc 4 - Knowledgeable - I have a technical understanding of the threats that exist online and what tools can be used to mitigate them
\emptycirc 5 - Expert - I conduct research or work in a field related to Internet privacy and security

\paragraph{VPN Usage}\mbox{}\\
\textit{In this section, we aim to understand the reasons why you use a VPN and your typical VPN usage patterns. Our definition of "VPN product" is any VPN service that you may have used for personal use. The service may have been free, on a trial basis, a subscription based service, or a one-time fee type service. We also include school/university provided VPNs in our definition. It does not include: Corporate (workplace specific) VPN configurations, or VPNs set up by an individual (such as using Algo, Outline, or Streisand)}\\\\
Q8. Have you ever used a commercial VPN service as defined above?* Examples include products like NordVPN, ExpressVPN; Free services like Hotspot Shield Free, TunnelBear Free, Psiphon, and Lantern; University VPNs provided to you (accessible using your university credentials).\\
\emptycirc Yes \emptycirc No\\\\
Q8A. What type of subscription do you generally use?\\
\emptycirc Free/trial version of a VPN service
\emptycirc Paid/premium version of a VPN service
\emptycirc Does not apply (University or Custom VPN)
\emptycirc Other (Please elaborate below)\\\\
Q9. Why do you use a VPN product? (Choose all that apply)\\
\emptysquare To access school or work networks remotely
\emptysquare To protect myself from various threats/adversaries
\emptysquare To access content blocked in my network (eg. due to censorship)
\emptysquare My IP address is blocked from certain websites
\emptysquare I'm interested in the technology behind it
\emptysquare To make public networks safer to use
\emptysquare For file sharing (e.g. torrents)
\emptysquare To access region-specific content e.g. on Netflix, Hulu, BritBox, News
\emptysquare It was a free service offered to me
\emptysquare Other (Please elaborate below)\\\\
Q10. How frequently do you use a VPN?\\
\emptycirc Occasionally
\emptycirc Every week
\emptycirc Every day
\emptycirc All the time/Always on
\emptycirc Other (Please elaborate below)\\\\
Q11. To make sure that you're still paying attention, please select only ``Yes, more than once" in the options below\\
\emptycirc Yes, once
\emptycirc Yes, more than once
\emptycirc No
\emptycirc I don't know

\paragraph{VPN Discovery and Choosing a VPN Product}\mbox{}\\
\textit{In this section, we aim to understand how you discover VPN products, and the decisions and trade-offs that you may have made while choosing a VPN provider.}\\\\
Q12. How difficult was it to decide which commercial VPN product/app to use?\\
\emptycirc Very difficult
\emptycirc Somewhat difficult
\emptycirc Neither easy nor difficult
\emptycirc Somewhat easy
\emptycirc Very easy\\\\
Q13. Please explain the reasons for your choice above.\\\\
Q14.  There are many resources that are aimed at helping users choose a VPN provider. We are interested in learning about all the resources that you used in your journey to select a VPN provider. What resources did you use to select your VPN provider? (Choose all that apply)\\
\emptysquare Actively researching on the Internet e.g. using a search engine
\emptysquare Recommendations from friends and family
\emptysquare Reading the VPN provider websites
\emptysquare Digital Training Workshops
\emptysquare I randomly encountered them while browsing the web, through advertisements
\emptysquare Recommendation websites e.g. TechRadar, CNET, Top10vpn
\emptysquare User review posts e.g. Reddit, YouTube
\emptysquare My work or school/university provides me a VPN
\emptysquare Conferences and events
\emptysquare Other (Please elaborate below)\\\\
\noindent\textit{The following question is displayed for each resource selected in Q14.}\\
Q14A.  In hindsight, how would you rate the level of credibility \& trustworthiness of the resources you selected for the previous question?\\
\textit{(Options available for each resource are \emptycirc Not Trustworthy \emptycirc No Opinion \emptycirc Moderately Trustworthy \emptycirc Extremely Trustworthy)}\\\\
Q15. Please rate the importance of the following criteria while selecting a VPN provider.*\\
\textit{(Options available for each criteria are \emptycirc Not preferred/Indifferent \emptycirc Preferred, but not a dealbreaker \emptycirc Required, 
a dealbreaker)}\\
$\rightarrow$ Speed (Quality of service)
$\rightarrow$ Price of the service
$\rightarrow$ Easy to understand/use app (GUI)
$\rightarrow$ Well-documented features
$\rightarrow$ Ability to change location to access media on websites e.g. Netflix, Hulu, or News
$\rightarrow$ Variety/number of servers located in different countries around the world
$\rightarrow$ Clear explanation of logging and data practices\\\\
\noindent\textit{The following question is displayed for each criteria selected in Q15.}\\
Q15A. Please move the tiles to rank the importance of the criteria you rated "Preferred" or "Required" in the previous question (1 being most important).*\\\\
Q16. Please select the names of all the commercial VPN providers you have tried/used before.\\
\emptysquare NordVPN
\emptysquare StrongVPN
\emptysquare ExpressVPN
\emptysquare Hide.me
\emptysquare IPVanish
\emptysquare Speedify
\emptysquare Hotspot Shield
\emptysquare Psiphon
\emptysquare TunnelBear
\emptysquare Calyx VPN
\emptysquare Windscribe
\emptysquare Mullvad VPN
\emptysquare Private Internet Access (PIA)
\emptysquare PrivateVPN
\emptysquare CyberGhost VPN
\emptysquare TorGuard
\emptysquare HideMyAss
\emptysquare Hola Free VPN
\emptysquare ProtonVPN
\emptysquare Astrill VPN
\emptysquare Norton Secure VPN
\emptysquare VyprVPN
\emptysquare SurfShark
\emptysquare Mozilla VPN
\emptysquare PureVPN
\emptysquare Others (please separate by commas)\\\\
Q17. Is there anything else about the process of VPN discovery that you wish to share with us?

\paragraph{Mental Model of VPNs and Your Personal Threat Model}
\textit{In this section, we ask questions about how you think VPNs work and your understanding of their data collection practices.}\\\\
Q18. How safe or unsafe do you feel when browsing the internet without a VPN?*\\
\emptycirc Very unsafe
\emptycirc Somewhat unsafe
\emptycirc No opinion
\emptycirc Somewhat safe
\emptycirc Very safe\\\\
Q18A. Please elaborate on why you feel that way when browsing without a VPN.\\\\
Q19. Imagine you are using a VPN and you open http://www.example.com, who do you believe knows that you have visited http://www.example.com? (Choose all that apply)\\
\emptysquare My Internet service provider (ISP)
\emptysquare My VPN provider
\emptysquare The owner of example.com
\emptysquare Nobody
\emptysquare I don’t know
\emptysquare Other (Please elaborate below)\\\\
Q20. Do you use a VPN for the purpose of securing or protecting your browsing activity?*\\
\emptycirc Yes
\emptycirc No\\\\
\textit{Display the following question if Q20: = Yes}\\
Q20A. Since you selected that you use a VPN to secure or conceal your browsing activity, who do you want to protect it from? (Choose all that apply)\\
\emptysquare My Internet service provider (ISP)
\emptysquare My School/Employer
\emptysquare Friends and family
\emptysquare Advertising companies
\emptysquare Hackers/Eavesdroppers on open WiFi networks
\emptysquare My government
\emptysquare Other governments
\emptysquare I do not use a VPN for this purpose
\emptysquare Other (Please elaborate below)\\\\
Q21.  To make sure that you're still paying attention, please select only "Somewhat safe" in the options below\\
\emptycirc Very unsafe
\emptycirc Somewhat unsafe
\emptycirc No opinion
\emptycirc Somewhat safe
\emptycirc Very safe\\\\
Q22. What data do you think is being collected about you by your VPN provider? (Choose all that apply)\\
\emptysquare My geolocation
\emptysquare Timestamps of when VPN is in use
\emptysquare VPN servers that I connect to
\emptysquare Websites visited
\emptysquare Interests/Preferences for ads
\emptysquare Demographics and account holder information
\emptysquare Device types
\emptysquare Private messages
\emptysquare Audio/Video collected from my device
\emptysquare Keystrokes recorded from my keyboard
\emptysquare I am not sure
\emptysquare Other (Please elaborate below and separate by commas)\\\\
Q23. Why do you think your VPN provider collects this data about you? (Choose all that apply)\\
\emptysquare Internal analytics and quality of service reasons
\emptysquare Advertising
\emptysquare User tracking
\emptysquare Political motives (government mandated)
\emptysquare Crime investigation (for law enforcement)
\emptysquare Selling information to third parties
\emptysquare I am not sure\\\\
Q24. Please use this text box to share your views on data collection by VPN providers\\\\
\paragraph{Expectations about VPN service}\mbox{}\\
\textit{In this section, we aim to discover what you expect from your VPN provider in terms of quality of service, privacy, and security. We aim to understand how VPN providers can build trust with you as a user.}\\\\
Q25. How safe or unsafe do you feel while using your favorite VPN as compared to browsing without the VPN?\\
\emptycirc Very unsafe
\emptycirc Somewhat unsafe
\emptycirc No opinion
\emptycirc Somewhat safe
\emptycirc Very safe\\\\
Q25A. Please elaborate on why you feel that way when browsing with your favorite VPN.\\\\
Q26. Please rate the level of concern you would have about the following VPN-related issues.\\
\textit{(Options available for each concern are \emptycirc Not at all concerned \emptycirc Slightly concerned \emptycirc Moderately concerned \emptycirc Very concerned \emptycirc Extremely concerned)}\\
$\rightarrow$ VPN provider logging my activity	
$\rightarrow$ VPN provider selling my activity data
$\rightarrow$ VPN servers’ geographic location not as advertised
$\rightarrow$ Lack of transparency or documentation
$\rightarrow$ VPN client software containing malware
$\rightarrow$ Misconfigured VPN servers leaking some data	
$\rightarrow$ VPN not communicating to me that the connection has dropped (VPN tunnel failure)\\\\
Q27. Which of these efforts would increase the trust you have towards a VPN provider?\\
\textit{(Options available for each effort are \emptycirc No Opinion \emptycirc Not at all important \emptycirc Slightly important \emptycirc Moderately important \emptycirc Very important \emptycirc Extremely important)}\\
$\rightarrow$ Independent Security Audits		
$\rightarrow$ Clear Logging Policy
$\rightarrow$ Response to Legal and Law Enforcement Requests (e.g. Warrant Canary)		
$\rightarrow$ Security Protocols and Disclosure of Breaches
$\rightarrow$ Endorsements from NGO and/or academics

\paragraph{End of Survey}\mbox{}\\
Q28.  Finally, is there anything related to commercial VPNs that you wish to share with us?

\section{VPN Provider Interview Questionnaire}
\label{app:provider-questionnaire}
\begin{enumerate}
  \item \textit{[Biggest Challenges]} Briefly tell us, what are the biggest problems that you see in the VPN ecosystem?
  \item \textit{[User Base]} Who is your main user base? There are many reasons why people use VPNs like circumvention, privacy from ISP etc, do you know why your users are using your VPN?
  \item \textit{[Features]} What features do you think users care about the most? What are the features you focus most development efforts on? 
  \item \textit{[Mental Models]} Do you think users have an accurate/good mental model of how VPNs work? Where do usability issues and frustration with VPNs usually come from?
  \item \textit{[Pricing]} From a user's perspective, we've also seen that pricing is a key criteria. How does the pricing work at your VPN? How does pricing in the industry work, generally?
  \item \textit{[Marketing/Recommenders]} Do you actively reach out to VPN recommenders to try your product? What strategies do you use to market your product?
  \item \textit{[Trust]} What are your efforts to build trust with users? Do you have third party security audits, why do you think they help? What are other efforts to build trust with users?
  \item \textit{[Dark Patterns]} What are some dark patterns and shady practices in the VPN ecosystem? 
\end{enumerate}

Miscellaneous (Extra questions, when we have time):
\begin{enumerate}[label=X\arabic*.]
    \item How do you think these issues can be fixed? How to combat these issues in the future?
    \item Are there technical challenges in implementation, like IPv6 for example? What are some of the barriers for these challenges?
    \item What are your bug disclosure models? 
    \item Do you have special circumvention technologies and obfuscation developed or implemented for censorship circumvention users?
    \item How do you reconcile with laws of the land where you have your servers?
\end{enumerate}
\end{document}